\documentclass[12pt]{article}
\usepackage[utf8]{inputenc}
\usepackage{pgf}
\usepackage{tikz}
\usepackage[numbers]{natbib}
\usetikzlibrary{positioning}
\usepackage{amsmath}
\usepackage{float}
\usepackage{graphicx}
\usepackage{tikz}
\usetikzlibrary{arrows}
\usepackage{caption}
\usepackage[export]{adjustbox}
\usepackage{lscape} 
\usepackage{multirow}
\usepackage{longtable}
\usepackage{nameref}
\usepackage[margin=1in]{geometry}
\usepackage{url}

\title{Triangulating Instrumental Variable, confounder adjustment and Difference-in-Difference methods for comparative effectiveness research in observational data}
\author{Laura G\"udemann, John M. Dennis, Andrew P. McGovern, \\ Lauren R. Rodgers,  Beverley M. Shields, William Henley \& Jack Bowden \\ on behalf of the MASTERMIND consortium}
\date{July 2022}

\begin{document}

\maketitle

\section*{Abstract}\label{sec:abstract}
Observational studies can play a useful role in assessing the comparative effectiveness of competing treatments. In a clinical trial the randomization of participants to treatment and control groups generally results in well-balanced groups with respect to possible confounders, which makes the analysis straightforward. However, when analysing observational data, the potential for unmeasured confounding makes comparing treatment effects much more challenging. Causal inference methods such as Instrumental Variable and Prior Even Rate Ratio approaches make it possible to circumvent the need to adjust for confounding factors that have not been measured in the data or measured with error. Direct confounder adjustment via multivariable regression and Propensity score matching also have considerable utility. Each method relies on a different set of assumptions and leverages different aspects of the data. 
\\
\\
In this paper, we describe the assumptions of each method and assess the impact of violating these assumptions in a simulation study. We propose the prior outcome augmented Instrumental Variable method that leverages data from before and after treatment initiation, and is robust to the violation of key assumptions. Finally, we propose the use of a heterogeneity statistic to decide if two or more estimates are statistically similar, taking into account their correlation. We illustrate our causal framework to assess the risk of genital infection in patients prescribed Sodium-glucose co-transporter-2 inhibitors versus Dipeptidyl peptidase-4 inhibitors as second-line treatment for Type 2 Diabets using  observational data from the Clinical Practice Research Datalink.
\\
\\
\noindent \textbf{Keywords:} causal inference, unmeasured confounding, triangulation, Instrumental Variable method, Prior Event Rate Ratio approach

\maketitle

\section{Introduction}\label{sec:introduction}
\begin{figure}[h]
	\centering
	\adjincludegraphics[height=4.7cm,trim={0cm 0cm 0cm 0cm},clip]{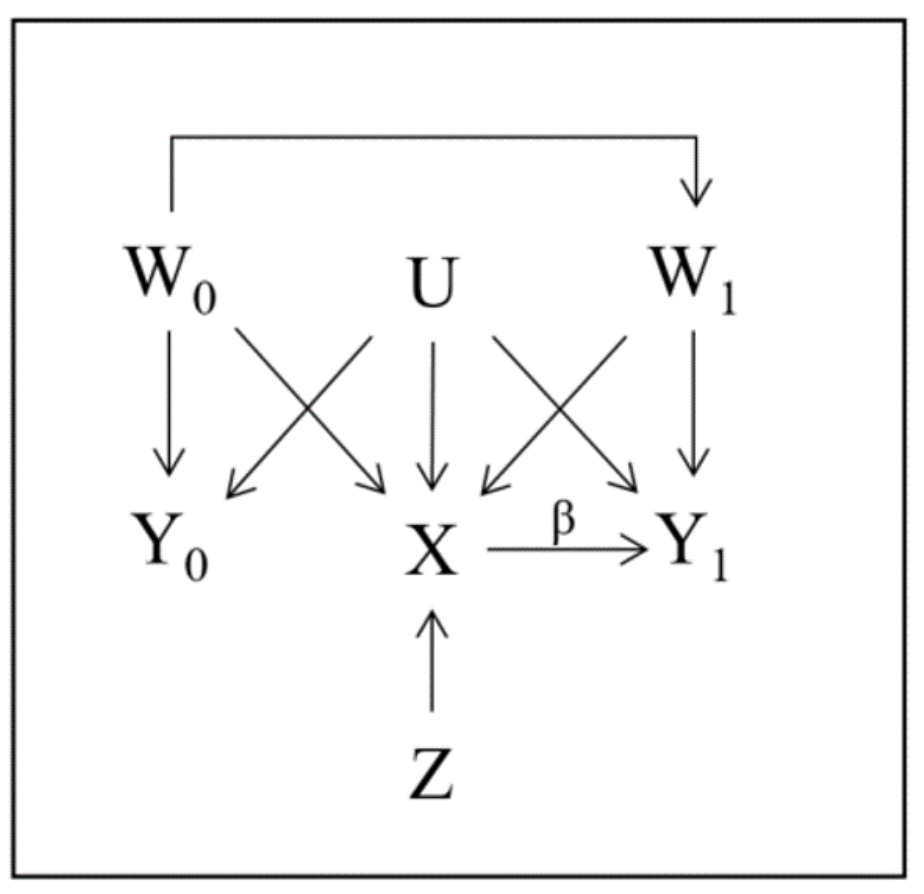}
	\caption{{ Causal diagram showing the relationship between $Y_{0}$, $X$, $Y_{1}$, $U$, $W_{0}$ and $W_{1}$ in the case where both the IV and DiD assumptions are satisfied. The estimates and assumptions are explained in detail in Section \ref{sec:methods}.}}
	\label{fig:basicDAGs}
\end{figure}

The gold standard approach for evaluating the efficacy of treatments is a randomized controlled trial (RCT). Due to strict specifications of RCTs with regard to  blinding and randomization of treatment assignment, causal conclusions about the treatment's effect on patient outcomes can be drawn without the need to adjust for prognostic factors, since they should be well-balanced across trial arms. This remains true even if the trial is affected by non-adherence, and non-adherence is predicted by the aforementioned prognostic variables. \cite{Greenland01}\\
\\
\\
\\
Observational data, for example from electronic healthcare records, provides vital means for assessing the comparative effectiveness of commonly prescribed medications with similar indications. Since these data are collected as part of routine care, treatment assignment is not randomized. This opens up the possibility that treatment choice (or the extent of treatment received) and patient outcomes may be simultaneously predicted by common variables, which could bias the analysis because of a lack of balance across treatment groups, in contrast to the adherence affected RCT setting previously discussed. This phenomenon is referred to as `confounding' and we will refer to such common factors  as confounders from now on. \cite{Greenland01, Jager08, Boyko13}
\\
\\
Standard causal inference methods such as stratification, multivariable regression or propensity score matching make it possible to analyse observational data and draw causal conclusions as long as all confounders can be accurately measured and appropriately adjusted for. \cite{Greenland01} For example, Dawwas et al. 2019 \cite{Dawwas19} used propensity score matched data for a retrospective cohort study for a comparative risk analysis of cardiovascular outcomes in people with Type 2 Diabetes (T2D) initiating Dipeptidyl peptidase-4 inhibitors (DPP4i) versus Sodium-glucose co-transporter-2 inhibitors (SGLT2i) therapy. Another example using standard causal inference methods is 
McGovern et al. 2020 \cite{McGovern20} who used multivariable Cox regression and propensity scores to define important clinical groups of people with T2D initiating either DPP4i or SGLT2i, with high risk of genital infection. \\
\\ 
Failure to measure and appropriately adjust for all confounders  can  bias  the estimation of the true  causal effect of treatment on the outcome of interest. Two causal inference approaches which circumvent the problem of unmeasured confounding are the Instrumental Variable (IV) and the Prior Event Rate Ratio (PERR) method. The IV approach addresses confounding by substituting each patients' observed treatment with a predicted treatment. This prediction is made using a variable that is assumed to be independent of any confounders and only affects the outcome through the treatment (the instrument). Randomization to a treatment group within a RCT is perhaps the best example of an IV, and can therefore be used to adjust for non-adherence. \cite{Greenland01,Streeter17, Bowden21b} Because of this, IV analyses using observational data are generally equated with the creation of a  pseudo-randomized controlled trial. Examples of IVs for observational data include geographic information such as the distance to the nearest health facility \cite{Penzzin15}, germ line genetic information \cite{Smith03}  or a physician's preference for a particular treatment \cite{Brookhart06b}. In this paper we will subsequently construct an IV of this latter type. 
\\
\\
The Prior Event Rate Ratio (PERR) method \cite{Rodgers20} is an alternative quasi-experimental approach which leverages data at two time points. Specifically, the outcome must be measured in the `prior' period before initiation of treatment  and then in the `study' period after treatment has commenced. The treatment effect is first estimated in the prior period by (somewhat paradoxically) regressing the prior outcome on the study period treatment indicator. This is assumed to capture the degree of unmeasured confounding in the treatment effect subsequently estimated in the study period, which can then be subtracted out to de-bias the analysis. The approach relies on the assumption of time invariant unmeasured confounding across both time periods. For related reasons it is necessary that a patient's outcome in the prior period does not influence the allocation of the study period treatment. Furthermore, the prior and study event of interest should be of the same nature and non-terminal, such as death.  \cite{Lin16, Rodgers20}. The PERR method is generally applied to time-to-event data, but is directly analogous to the method of Difference-in-Difference (DiD) regression in the case of continuous or binary outcomes.    
\\
\\
Figure \ref{fig:basicDAGs} shows a causal diagram illustrating the possible relationship between: the outcome in the prior and study periods ($Y_{0}$ and $Y_1$ respectively); the treatment indicator ($X$); an IV  ($Z$); measured confounders of treatment and outcome in both periods ($W_{0}$ and $W_{1}$ respectively); and unmeasured confounders ($U$). In this diagram, the assumptions of both the DiD (analogous to PERR) and IV approaches are satisfied, but the `no unmeasured confounder' (NUC) assumption underlying a direct adjustment strategy is not.

\noindent In this paper we consider the joint application of direct confounder adjustment, IV and DiD approaches for estimating the causal effect of treatment using observational data. In Section \ref{sec:methods} we give a more detailed description of each method and introduce a heterogeneity statistic to decide if two or more estimators are sufficiently similar. In Section \ref{sec:sim1} we assess the performance of these approaches in a detailed simulation study.
In Section \ref{sec:exIV} we consider an extension of the standard IV approach using pre- and post-treatment outcome data that can be used in scenarios where the assumptions of both the standard IV and DiD approaches are violated. We call this extension the prior outcome augmented Instrumental Variable method (POA-IV). In Section \ref{sec:application} we apply our methods to routinely collected health care records to assess the causal effect of SGLT2i compared to DPP4i as second-line therapy on the risk of genital infections, exploiting variation in general practitioner prescribing habits to construct an IV. We conclude in Section \ref{sec:discussion} with a discussion and point to further research. Source code for this research for all simulations and the application study in this paper is available at \url{https://github.com/GuedemannLaura/POA-IV}.

\section{Methods}\label{sec:methods}
We are interested in estimating the comparative effectiveness of two treatments ($X = 1$ compared to $X = 0$) on outcome $Y_1$ using observational data. Using the potential outcome notation, let $Y_{1i}(X_i = x)$ denote the outcome of patient $i$ if assigned treatment level $X_i = x$. The target of this analysis is a hypothetical estimand: 
\begin{equation}\label{equ:causal_estimand}
    \beta = E[Y_{1i}(X_i = 1)] - E[Y_{1i}(X_i = 0)]
\end{equation}
That is, the difference in expected outcomes if all patients could receive treatment level 1 compared to treatment level 0. For simplicity, we will assume in this section that the outcome of interest is continuous, with the extension to binary outcomes discussed in Section \ref{sec:binary_outcome}.

\subsection{The `As Treated' and `Corrected as Treated' estimate}
In a RCT with complete adherence to the assigned treatment, hypothetical estimand $\beta$ could be consistently estimated using the `as Treated' estimate, by comparing the average outcome across both treatment groups: 
\begin{equation}\label{equ:as_treated_estimate}
    \hat{\beta}_{aT} = \hat{E}[Y_1|X=1] - \hat{E}[Y_1|X=0] \nonumber .
\end{equation}
Complete adherence could be illustrated in Figure \ref{fig:basicDAGs} by letting $Z$ represent the randomized treatment assignment and removing all arrows into $X$ from $W_{0}$, $W_{1}$ and $U$, so that only the $Z\rightarrow X$ arrow remains. Difficulties emerge when calculating the as treated estimate with observational data, because treatment assignment is not randomized or controlled by the researcher. It is then possible that factors exist which simultaneously affect (or confound) the treatment assignment and the outcome.  This would lead to an imbalance across the treatment groups with respect to $W_{0}$, $W_{1}$ and $U$  and the estimate $\hat{\beta}_{aT}$ will consequently be biased due to confounding. \\
\\
If all confounding factors are known and can be appropriately measured and adjusted for - which we call the `no unmeasured confounder' (NUC) assumption -  a `Corrected as Treated' (CaT) estimate that is additionally adjusted for these factors can consistently estimate  $\beta$. Returning to Figure \ref{fig:basicDAGs}, if the NUC assumption held so that $U$ was absent from the diagram it would be sufficient to adjust for  $W_{1}$ since $W_{0}$ only affects $Y_{1}$ through $W_{1}$ and the CaT estimate would be
\begin{equation}\label{equ:corrected_as_treated_estimate}
    \hat{\beta}_{CaT}= \hat{E}[Y_1|X=1, W_1] - \hat{E}[Y_1|X=0, W_1]. 
\end{equation}
This could be estimated from fitting the following multivariable regression of $Y_{1}$ on $X$ and $W_{1}$  as

\begin{equation}\label{equ:CaT_model}
    E[Y_1|X,W_1] = \beta_0 + \beta_{CaT}X + \beta_1 W_1 \nonumber.
\end{equation}

\subsection{The Instrumental Variable estimate}\label{sec:IV}
In many settings the NUC assumption may be thought unreasonably strong. The Instrumental Variable (IV) method offers a means for circumventing the problem of unmeasured confounding to consistently estimate the hypothetical estimand. It works via the construction of a pseudo-randomized controlled trial using a variable $Z$ which needs to fulfill the following three assumptions in order to be a valid IV:
\begin{itemize}
    \item{IV1: $Z$ is associated, or predicts $X$;}
    \item{IV2: $Z$ is independent of $Y_1$ given $X$ and $U$;}
    \item{IV3: $Z$ and $Y_1$ do not share a common cause.}
\end{itemize}
IV1 is often referred to as the relevance assumption and the $Z-X$ relationship can be empirically tested from a regression of $X$ on $Z$. The assumption would be invalidated if this association is weak, with an $F$-statistic of at least 10 often used as a threshold for good strength of an IV. \cite{Stock02} Assumption IV2 is also referred to as the exclusion restriction and requires $Z$ to only influence $Y_1$ through $X$ but not directly. IV3, the exchangeability assumption, requires that $Z$ and $Y_1$ are not themselves confounded. \cite{Greenland01, Aso20, Lousdal18}.\\
\\
In its simplest form where only one IV is used and no adjustment for covariates is made, the IV estimate for $\beta$ is the ratio of the $Y_{1}-Z$ association and the $X-Z$ association: 
\begin{equation}
    \hat{\beta}_{IV} = \frac{\hat{E}[Y_1|Z=1] - \hat{E}[Y_1|Z=0]}{\hat{E}[X|Z=1] - \hat{E}[X|Z=0]} \quad. 
\end{equation}
In order to enable consistent estimation of hypothetical estimand (\ref{equ:causal_estimand}) using (3), we additionally make the homogeneity assumption that the average treatment effect is constant across both levels of the IV $Z$, at each level of the treatment \cite{Bowden21b}:
\[\
E[Y_{i}(X=1)-Y_{i}(X=0)|Z=1,X=x]=E[Y(X=1)-Y(X=0)|Z=0,X=x].
\]

\noindent A more general method for IV estimation with a continuous outcome that allows for multiple IVs and covariate adjustment is Two-Stage Least Squares (TSLS) \cite{Aso20}. To implement TSLS with a single IV $Z$ and the measured confounder $W_{1}$, we first fit a logistic regression model for $X$ given $Z$ and $W_{1}$:
\begin{equation}\label{equ:IV_1}
    \text{Logit}(P(X=1|Z, W_1)) = \alpha_{X,0} + \alpha_{X,Z}Z + \alpha_{X,W_1}W_1
\end{equation}
\noindent The estimated coefficients of this model are then used to predict $X$ given $Z$ and $W_{1}$  as $\hat{X}$, which is then itself regressed on $Y_{1}$ and $W_{1}$ in a second-stage model:
\begin{equation}\label{equ:IV_2}
    E(Y_1|\hat{X}, W_1) = \alpha_{Y_1,0} +  \beta_{IV}\hat{X} + \alpha_{Y_1,W_1}W_1 
\end{equation}
The coefficient of $\hat{X}$ is then taken as the TSLS estimate \cite{Lousdal18}. Using a valid IV, the TSLS is consistent under the homogeneity assumption and additionally that the covariates are correctly modelled in (5) \cite{Wooldridge19, Bowden84}.

\subsection{Difference-in-Difference estimate}\label{sec:DiD}
An alternative approach to adjust for unmeasured  confounding is the difference-in-difference (DiD) estimate. It can be applied to continuous and binary outcomes and is conceptually equivalent to the Prior Event Rate Ratio (PERR) method typically applied to time-to-event outcomes \cite{Lin16,Rodgers20}. Borrowing the terminology of the PERR approach,  DiD estimation leverages data from two periods: the {\it prior} period before drug initiation and the {\it study} period after drug initiation. For the estimation of the treatment effect in the study period, the treatment effect measured for the prior period is used to capture the degree of unmeasured confounding. The method presumes that the treatment effect measured in the prior period reflects the composite effect of measured and unmeasured confounders on the outcome, if none of the participants receive any of the study treatments in the prior period. \cite{Weiner08, Tannen08} Once estimated, it can then be subtracted from the $X-Y_{1}$ association (or the as Treated estimate). This approach relies on the following assumptions: 
\begin{itemize}
    \item DiD1: $Y_0$ does not influence the treatment decision $X$ directly 
    \item DiD2: The effect of $U$ on the outcome is constant across time conditional on $W_0$ and $W_1$. \citep{Lin16, Rodgers20}
\end{itemize}
Previous studies show that the DiD method is biased in case of the violation of assumption DiD1  \cite{Uddin15, Gallagher09} and DiD2 \cite{Yu12}. A formal proof that these assumptions are sufficient for identification of hypothetical estimand (\ref{equ:causal_estimand}) is given in Appendix 2. Both assumptions are satisfied in Figure \ref{fig:basicDAGs}.
\noindent For continuous outcomes the DiD estimate can be calculated by subtracting the results of two linear regressions from the prior and study period:
%\begin{equation}
%\begin{split}
%   E[Y_1 - Y_0|X, W_0, W_1] 
%   & = \gamma_{Y_1,0} + %\betaX + \gamma_{Y_1,W_1}W_1 
 %   - (\gamma_{Y_0,0} + %\betaX + %\gamma_{Y_0,W_0}W_0) \\
 %  & = \gamma_{Y_1,0} - %\gamma_{Y_0,0}
 %  + \underbrace{(\beta - %\beta)}_\text{$\beta_{DiD}$}X + \gamma_{Y_1,W_1}W_1 - \gamma_{Y_0,W_0}W_0 
%\end{split}
%\end{equation}

%\noindent and the estimate will be 
\begin{equation}\label{equ:DiD_estimate}
\begin{split}
\hat{\beta}_{DiD} & = \hat{E}[Y_1|X=1, W_1] - \hat{E}[Y_1|X=0, W_1]  \\
 & -(\hat{E}[Y_0|X=1, W_0] - \hat{E}[Y_0|X=0, W_0]) . 
\end{split}
\end{equation}
The DiD estimate can also be calculated for a sample of $n$ individuals via the following single regression model: 
\begin{equation}\label{equ:DiD_regresion}
    E[Y|X^*,W,P] = \gamma_0 + \gamma_PP + \gamma_{X^*}X^* + \beta_{DiD} P\cdot X^* + \gamma_w W + \gamma_{WP} W \cdot P.
\end{equation}
\noindent Here, $X^* \in \{0,1\}$ and $P \in \{0,1\}$ are $2n$-length treatment and period indicator variables, $Y=(Y_{0},Y_{1})$, $W=(W_{0},W_{1})$ summarize the information of outcomes and covariates for both periods.The regression coefficients of the $P\cdot X^*$ interaction term is taken as the DiD estimate. \cite{Zhou16} Fitting this model facilitates the easy extraction of a standard error for the DiD estimate directly from the hessian matrix.\\
\\
The DiD method utilizes only two outcome measurements before and after treatment initiation  and can be viewed as a simple special case of an interrupted time series analysis, which incorporates data from multiple time points within a formal longitudinal model \cite{Bernal19, Craig12, Rockers15, Soumerai15}. Due to the limitation of our own data on outcome measurements and our focus on triangulating findings across methods, we restrict our attention to the DiD approach in this paper.

\subsection{Extension to binary outcomes}\label{sec:binary_outcome}
In case of a binary outcome the comparative treatment effect can be estimated using the CaT, IV, CF and DiD models using logit or probit models, instead of linear regressions for continuous outcomes. Whilst TSLS is the standard tool for IV analysis with continuous outcomes,  the control function (CF) method is typically used for binary outcomes. This method can accommodate linear and non-linear associations between the IV and treatment and between the treatment and the outcome \cite{Aso20, Techetgen14}.   For the first stage model and continuous outcomes the IV is regressed on treatment and all measured confounders, as shown in model \ref{equ:IV_1}. From this regression the residual $\hat{\Delta} = X - E(X| Z, W_1)$ is calculated and used in the second stage model with
\begin{equation}
    \text{Logit}(P(Y_1=1|X, Z, W_1, \hat{\Delta})) = \beta_{Y_1,0} + \beta_{Y_1,W_1} W_1 + \beta_{CF} X + (\beta_{Y_1, \hat{\Delta}} +  \beta_{Y_1, Z \hat{\Delta}}Z) \hat{\Delta} . 
\end{equation}
We will use the both the standard IV and CF approaches for IV analyses going forward.
\\
\\
When employing  logistic regression models for the CaT, IV, CF and DiD models, for our purposes we prefer to extract the treatment estimate as a risk difference, or an Average Marginal Effect (AME) \cite{Bowden21b}. For example, in the case of the CF estimate, after fitting model (8) to obtain estimates for its constituent parameters, the AME is calculated as the difference in average predicted probabilities when $X$ is fixed at 1 and 0 respectively:

\begin{equation}
\hat{\beta}_{CF} = \frac{1}{n}\sum^{n}_{i=1}\left\{\hat{P}(Y_{1}=1|X=1,Z,W,\hat{\Delta})-\hat{P}(Y_{1}=1|X=0,Z,W,\hat{\Delta})\right\} \nonumber    
\end{equation}
In R, this can easily be done using the {\tt  margins()} package \cite{Leeper21}. Choosing a scale that is collapsible makes it more straightforward to compare estimates across different methods which use different covariate adjustment sets. Even if their respective assumptions are all satisfied, using a non-collapsible scale such as an odds ratio could mean that the underlying causal estimands of two methods are in fact distinct \cite{Huitfeldt19}.

\subsection{Similarity statistics}
In order to assess the similarity of the estimates, after taking care to estimate them on the same scale and whilst accounting for their correlation, we use a generalized heterogeneity statistic \cite{Slotani64} of Cochran's Q \cite{Cochran54} of the form

\begin{equation}
  Q_e = (\boldsymbol{\hat{\beta}}_e - \hat{\beta}_{IVW,e})\hat{\Sigma}^{-1}_{e}(\boldsymbol{\hat{\beta}}_e - \hat{\beta}_{IVW,e})^{T} \label{eq:QSimilarity}
\end{equation}
where
\begin{itemize}
    \item{$e$ is the set of estimates, for example $\{CaT, IV, DiD\}$;}
    \item{$\boldsymbol{\hat{\beta}}_e$ is a vector of all estimates in $e$ with $j$th entry $\hat{\beta}_{ej}$;}
    \item{$\hat{\beta}_{IVW,e}$ is the inverse variance weighted average of all estimates in $e$;}
\item{$\hat{\Sigma}_{e}$ is the covariance matrix for $\boldsymbol{\hat{\beta}}_e$, approximated by a non-parametric bootstrap;}    
\item{
$\hat{\beta}_{IVW,e}$ is calculated using $w_{ej}$ the inverse variance of the corresponding estimate and 
\begin{equation*}
    \hat{\beta}_{e} = \frac{\sum_{j \in e} w_{ej}  \hat{\beta}_{ej}}{\sum_{j \in e} w_{ej}}.
\end{equation*}}
\end{itemize}
Under the assumption that all estimates in $e$ are targeting the same underlying quantity, $Q_{e}$ is asymptotically $\chi^{2}_{e-1}$ distributed. This assumption is rejected at level $\alpha$ if  $Q_e > \chi^2_{e-1}(1-\alpha)$ and the estimates are assumed not to be similar. Equation (\ref{eq:QSimilarity}) can therefore be used to assess the extent of agreement across estimators and, by extension, the validity of the assumptions that they rest on. This approach can be seen as a generalisation of the causal triangulation framework for uncorrelated estimates described in Bowden et. al. \cite{Bowden21a}. We showcase an application of the $Q_e$ in Section \ref{sec:Q_e}.

\section{Simulation Study}\label{sec:sim1}

In this section we use Monte-Carlo simulation to study the performance of  the CaT, IV (TSLS and CF) and DiD estimates in scenarios where their specific assumptions are variously satisfied and violated. The simulation was conducted in R Studio (version 4.1.2)  and the set-up is motivated to a degree by the applied analysis in Section \ref{sec:application}.

\subsection{Simulation set up}
\begin{landscape}
\begin{figure}
	\centering
	\includegraphics[width=.6\linewidth, angle =-90, left]{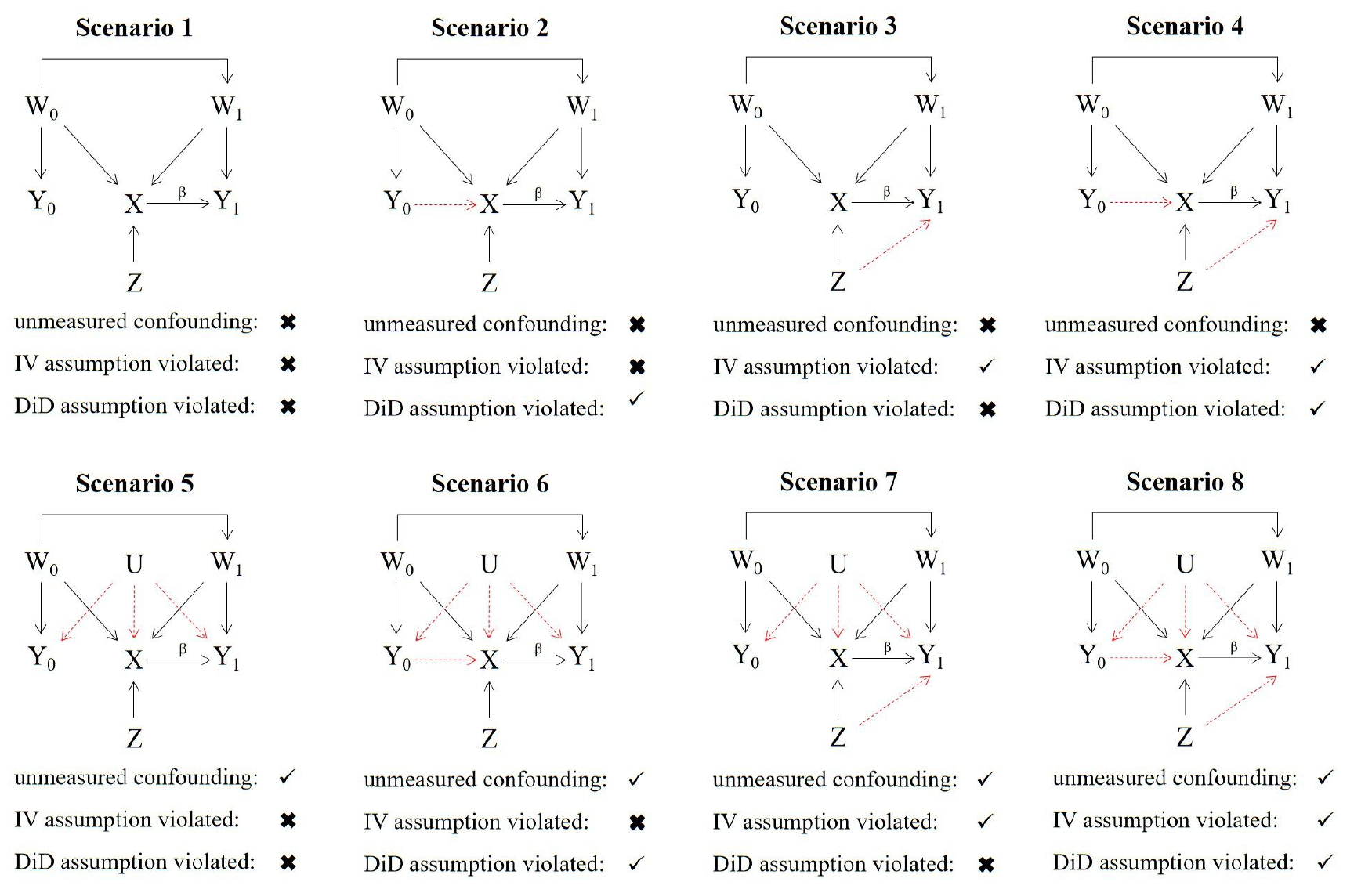}
	\caption{DAGs representing the scenarios of the simulation.}
	\label{fig:scenariossimuation1}
\end{figure}
\end{landscape}

Across 1000 independent simulations, observational data is generated for n = 5000 patients grouped into $n_{g} = 50$ clusters, with each cluster representing a general practitioner practice. The full data generating models are summarized in Appendix 1 but the main features are now described. The treatment group indicator $X$ and the outcome variables, $Y_0$ and $Y_1$ are simulated as binary variables, in each case representing the presence or absence of a binary adverse event. The true treatment effect, quantified on the risk difference scale, $\beta = 0.1$. Therefore, the average causal effect of treatment 1 versus 0 is a $10\%$ increase in adverse event risk. Further information on the chosen parameter values can be found  in the additional provided material at \url{https://github.com/GuedemannLaura/POA-IV}.\\ 
\\
Treatment and outcome variables are allowed to depend in principle on: measured confounders, $W_0$ and $W_1$; one unmeasured confounder $U$ (normally distributed); and the IV $Z$ (simulated as a binary variable). Specifically, $Z$ is constant at the practice level, and therefore conveys information about practitioners' preference to prescribe one treatment over the other.  
\\
\\
Figure \ref{fig:scenariossimuation1} summarizes the 8 scenarios implemented in the simulation. 
In scenario 1, the NUC, IV and DiD assumptions are all satisfied. In scenarios 2-4, the NUC assumption is satisfied, but certain IV and DiD assumptions are violated. Specifically, in Scenario 2, the DiD1 assumption which requires $X$ to be independent of $Y_0$ is violated. This could be, for example, because the occurrence of an adverse event in the prior period influences the practitioners' treatment decision in the study period. In scenario 3, IV2 (exclusion restriction) is violated. This could for example represent the case where practitioners' preference for treatment is associated with their tendency to record adverse events. Previous knowledge about adverse events for a specific treatment could easily give rise to this effect. Both the IV2 and the DiD1 assumptions are violated simultaneously in scenario 4. In scenarios 5 to 8 unmeasured confounding is present (NUC violated) although it is constant over the two time periods. In addition to the unmeasured confounding,  the DiD1 and IV2 assumptions are violated in scenarios 6 and 7 respectively. In Scenario 8 the NUC, IV2 and DiD1 assumptions are all violated. The eight scenarios are illustrated using causal diagrams in Figure 2. The CaT, IV, CF and DiD estimates were calculated by fitting the models listed in Table 1 using logistic regression in tandem with the {\tt  margins()} package \cite{Leeper21}, as described in the previous section. As IV2 is violated in scenarios 3, 4, 7 and 8, $Z$ becomes a measured confounder of $Y_1$. Is it therefore included in the CaT and DiD model for these scenarios. 

\begin{table}[h]
\centering
\begin{tabular}{l l|l}
 \hline
 \multicolumn{2}{c|}{Estimate}    & Fit                 \\ \hline
CaT & $\hat{\beta}_{CaT}$       &      $Y_1 \sim X + W_1$               \\
&    &   \\                  
IV &  $\hat{\beta}_{IV}$ &   First stage model: $X \sim W_1+ Z $\\
&    & Second stage model: $Y_1 \sim \hat{X} + W_1$  \\
&    &   \\
CF &  $\hat{\beta}_{CF}$ &   First stage model: $X \sim W_1 + Z $\\
&    & Second stage model: $Y_1 \sim X + W_1 + \hat{\Delta} + \hat{\Delta} \cdot Z$   \\
&    &   \\
DiD &  $\hat{\beta}_{DiD}$  &  $Y \sim P + X^* + P \cdot X^* + W + W \cdot P$ \\
&    &   
\end{tabular}
\caption{\label{tab:model_fit_sim1} { Summary of the models for CaT, IV, CF and DiD fitted in the simulation. For scenarios 3, 4, 7, 8 $Z$ is included in the DiD and CaT model as measured confounder.}}
\end{table}

\subsection{Simulation results}
\begin{figure}
	\centering
	\includegraphics[width=0.8\linewidth]{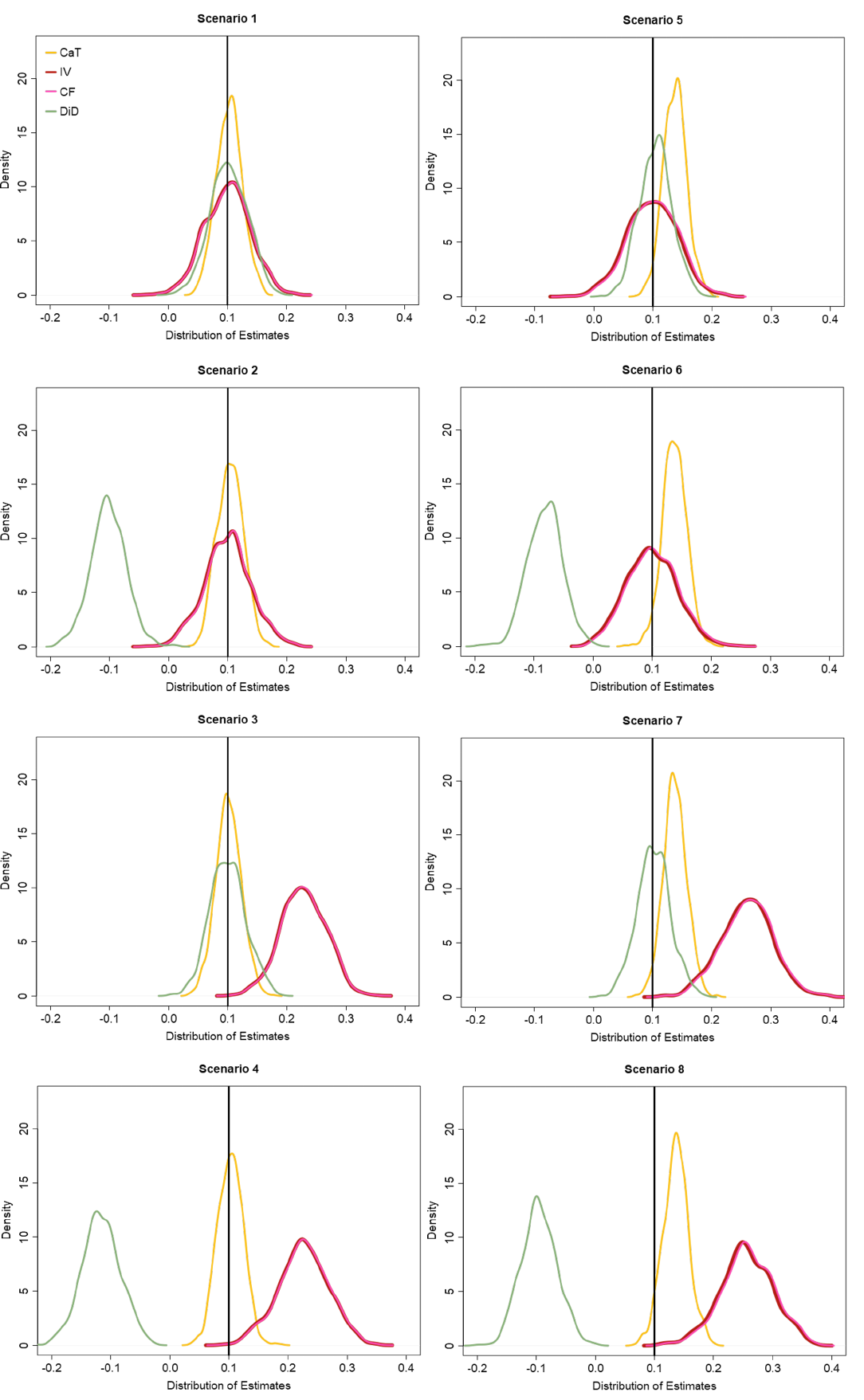}
	\caption{{ Distribution of estimation results for the CaT, IV, CF and DiD method for all simulation scenarios.}}
	\label{fig:results_simulation1}
\end{figure}

Simulation results are summarized for all 8 scenarios in Table  \ref{tab:results_sim1}. Specifically, we use the 1000 CaT, IV, CF and DiD estimates to calculate the: bias, standard deviation (SD) and mean squared error (MSE); the mean empirical standard error (SE) arising directly from the model fits; coverage rate of 95\% confidence intervals and the type I error (T1E) rate when rejecting the null hypothesis of no causal effect at the 5\% significance level. In order to assess the type 1 error simulation calculations were executed with $\beta = 0$.  The most efficient and unbiased method for each scenario is highlighted in Table \ref{tab:results_sim1} in {\bf bold}. Figure \ref{fig:results_simulation1} shows the distribution of the CaT, IV, CF and DiD estimates over all simulation runs. Additional simulation results including the Monte Carlo standard error estimates of the performance measures as described in Morris et al. 2018 \cite{Morris19}, are given in Appendix 3.
\\
\\ 
\noindent For scenario 1, we confirm that the CaT, IV, CF and DiD estimates are all unbiased for the hypothetical estimand $\beta$, and the CaT estimate is most efficient. The coverage and T1E rates for all estimates are close to their nominal levels. In scenario 2, the DiD estimate is systematically biased and consequently has poor coverage and T1E. Since a non-zero $Y_{0}-X$ relationship does not affect the CaT, IV or CF approaches, they estimate the treatment effect without bias, with the CaT being the most efficient. In scenario 3, which was intended to showcase the impact of IV2 assumption violation only, the IV and CF estimates are biased and their coverage/T1E rates are also adversely affected. The diagram for scenario 3 in Figure \ref{fig:scenariossimuation1} reveals that due to the direct effect of $Z$ on $Y_1$, $Z$ is an additional confounder, which must be included in the regression models for the CaT and DiD estimates. The results for the CaT and DiD estimates therefore include $Z$ as measured confounder. Again, the CaT estimate is the most efficient in this scenario. In scenario 4 both the IV and DiD assumptions are violated and $Z$ becomes a measured confounder for $Y_1$ again. Results of the CaT and DiD estimates additionally adjusted for $Z$ as a confounder show that only CaT remains unbiased in this scenario.  \\
\\
\noindent For scenario 5 to 8, unmeasured confounding was implemented. Consequently, the CaT estimate is biased and displays lower coverage and higher T1E compared to the IV, CF and DiD estimates, which work as planned.  Comparing the latter two, the DiD estimate is the most efficient unbiased effect measure in scenario 5. When the DiD1 assumption is also violated in scenario 6, its estimate is again biased and shows very low coverage and high T1E rates. Only the IV and CF estimates remain unbiased. Similar to scenario 4, in scenario 7 the DiD estimate will be only biased if $Z$ is not included in the model. The CaT estimate on the other hand remains biased due to the unmeasured confounding.
In scenario 8, the identifying assumptions of all three methods are violated. Consequently, none of the methods are able to estimate the treatment effect without bias. This scenario may very well represent the reality of a given analysis setting. For this reason in Section \ref{sec:exIV} we discuss an extension of the standard Instrumental Variable method  that can give consistent estimates for the causal effect under a different set of assumptions. 

%--------------------------------------------------------------------------------

\newpage
\begin{longtable}{l|lllll}
\hline
\multicolumn{1}{c|}{  } &   & CaT     & \multicolumn{1}{c}{IV} & \multicolumn{1}{c}{CF}  & \multicolumn{1}{c}{DiD} \\ \hline
\endhead
\multirow{5}{*}{Scenario 1}    & bias & \textbf{0.313} & -0.0117  & 0.0616 & 0.251 \\ 
 & S.E. & \textbf{0.0685} & 0.121 & 0.121 & 0.0988 \\ 
 & MSE & \textbf{0.0478} & 0.146 & 0.146 & 0.0981 \\ 
 & coverage & \textbf{94.1} & 95.2 & 95 & 93.4 \\ 
 & T1E & \textbf{6} & 5.7 & 5.5 & 5 \\    \hline
\multirow{5}{*}{Scenario 2}    & bias & \textbf{0.411} & -0.0345 & 0.0951 & -20.2 \\ 
  & S.E. & \textbf{0.0697} & 0.125 & 0.125 & 0.0956 \\ 
  & MSE & \textbf{0.0502} & 0.156 & 0.157 & 4.17 \\ 
  & coverage & \textbf{95} & 95.2 & 95.2 & 0 \\ 
  & T1E & \textbf{5.3} & 4.8 & 4.8 & 100 \\     \hline
\multirow{5}{*}{Scenario 3}    & bias & \textbf{0.134} & 12.8 & 12.9 & -0.002 \\ 
  & S.E. & \textbf{0.0687} & 0.121 & 0.121 & 0.0958 \\ 
  & MSE & \textbf{0.0473} & 1.78 & 1.8 & 0.0916 \\ 
  & coverage & \textbf{94.2} & 8.1 & 8 & 94.3 \\ 
  & T1E & \textbf{4.6} & 94.9 & 94.8 & 6.3 \\ \hline
\multirow{5}{*}{Scenario 4}    & bias & \textbf{0.261} & 12.6 & 12.7 & -21.6 \\ 
  & S.E. & \textbf{0.0685} & 0.135 & 0.135 & 0.102 \\ 
  & MSE & \textbf{0.0475} & 1.77 & 1.8 & 4.78 \\ 
  & coverage & \textbf{96.4} & 13.9 & 12.9 & 0 \\ 
  & T1E & \textbf{4.7} & 88.3 & 88.4 & 100 \\     \hline
\multirow{5}{*}{Scenario 5}    & bias & 3.72 & -0.169 & 0.111 & \textbf{0.436} \\ 
  & S.E. & 0.0625 & 0.134 & 0.133 & \textbf{0.0868} \\ 
  & MSE & 0.177 & 0.178 & 0.177 & \textbf{0.0771} \\ 
  & coverage & 53.4 & 95.2 & 94.7 & \textbf{95} \\ 
  & T1E & 45.4 & 5.9 & 5.9 & \textbf{4.8} \\     \hline
\multirow{5}{*}{Scenario 6}  &  bias & 3.77 & \textbf{-0.16} & \textbf{0.184} & -18.2 \\ 
  & S.E. & 0.0649 & \textbf{0.135} & \textbf{0.135} & 0.0935 \\ 
  & MSE & 0.184 & \textbf{0.182} & \textbf{0.182} & 3.39 \\ 
  & coverage & 55.4 & \textbf{96.4} & \textbf{96} & 0 \\ 
  & T1E & 38.5 & \textbf{5.6} & \textbf{5.6} & 100 \\    \hline
  %\pagebreak
\multirow{5}{*}{Scenario 7}    & bias & 3.8 & 15.9 & 16.2 & \textbf{0.205} \\ 
  & S.E. & 0.0631 & 0.14 & 0.14 & \textbf{0.0894} \\ 
  & MSE & 0.184 & 2.72 & 2.82 & \textbf{0.0803} \\ 
  & coverage & 55.5 & 6.2 & 5.2 & \textbf{94.6} \\ 
  & T1E & 45.3 & 96.5 & 96.8 & \textbf{5.1} \\   \hline
\multirow{5}{*}{Scenario 8}    & bias & 3.6 & 15.5 & 15.9 & -19.5 \\ 
  & S.E. & 0.0682 & 0.141 & 0.141 & 0.0958 \\ 
  & MSE & 0.176 & 2.61 & 2.72 & 3.9 \\ 
  & coverage & 60.4 & 7.4 & 6.5 & 0 \\ 
  & T1E & 39.4 & 94.8 & 95.1 & 100 \\                
\caption{ Bias, standard errors (SE) and mean squared error (MSE) (all $\times$ 100);  coverage and type 1 error (T1E) rate (both expressed as a percentage based on a 95\% confidence interval and 5\% significance threshold) for the estimates CaT, IV, CF and DiD and for all scenarios. }\label{tab:results_sim1}
\end{longtable}

%---------------------------------------------------------------------------------------------------------------------------------

\subsection{Similarity statistic performance}\label{sec:Q_e}
As proof of concept for the $Q_e$ statistics explained in Section \ref{sec:methods}, we repeat simulation scenario 1, 2, 3 and 5 with 500 simulation runs. In each simulation run, $Q_e$ is calculated using $500$ non-parametric bootstraps. Table \ref{tab:Sim_Q_statistics_results} shows the rejection rates (in \%) when testing if the CaT, CF and DiD estimates are similar at the 5\% level using all three pairwise comparisons. 

\begin{table}[h]\centering
\begin{tabular}{l|l|ll}
\hline
 &\multicolumn{1}{c|}{${e}$} & \multicolumn{1}{c}{$H_0$ rejected} & \multicolumn{1}{c}{$95\%$ CI}       \\ \hline
\multirow{3}{*}{Scenario 1} 
& CaT, CF      & 7.4     & 5.11; 9.7   \\
& CaT, DiD     & 6.4     & 4.25; 8.55   \\
& CF, DiD      & 6.4     & 4.25; 8.55 \\
\hline
\multirow{3}{*}{Scenario 2} 
& CaT, CF      & 6.0    &  3.92; 8.08  \\
& CaT, DiD     & 100.0     &  100; 100   \\
& CF, DiD      & 100.0     & 100; 100 \\
\hline
\multirow{3}{*}{Scenario 3} 
& CaT, CF      & 96.2    & 94.52; 97.88   \\
& CaT, DiD     & 4.4     &  2.6; 6.2   \\
& CF, DiD      & 87.8     &  84.93; 90.67\\
\hline
\multirow{3}{*}{Scenario 5} 
& CaT, CF      & 18.2    &  14.82; 21.58    \\
& CaT, DiD     & 40.8     &   36.49; 45.11  \\
& CF, DiD      & 4.0     &  2.28; 5.72
\end{tabular}
\caption{\label{tab:Sim_Q_statistics_results} { Rejection rates and $95\%$ confidence intervals (in $\%$). Results are shown for all pairwise combinations of the estimates.}}
\end{table}
\noindent In Scenario 1 all three estimators  target the same true estimand $\beta$. Although we see a small degree of type I error inflation, rejection rates for each test are reassuringly low. In scenario 2 the data is generated without unmeasured confounding, but the DiD1 assumption is violated. Therefore the DiD estimator targets a distinct estimand from the CaT or CF approaches, which themselves target the same  estimand. This is indeed reflected by the test results, where we observe 100\% power to detect a difference between the DiD estimate and either the CaT or CF estimates and a low power of 6\% to distinguish the CaT and CF estimates themselves. In Scenarios 3 (where only the CaT and DiD estimates are truly similar) and Scenario 5 (where only CF and DiD estimates are truly similar) the $Q_{e}$ statistic exhibits comparable performance. In scenario 5 we can only reliably detect that the DiD and CF estimates are similar but we do not have enough power to detect the differences with the CaT estimate. This might be because the bias due the violation of the NUC assumption is relatively small as well as the true difference in estimates. 

\section{The prior outcome augmented Instrumental Variable method}\label{sec:exIV}
In this section we introduce the prior outcome augmented Instrumental Variable (POA-IV) estimate which aims to overcome the limitations of the DiD and standard IV estimate by leveraging data from both the prior and study period. Specifically, we look to leverage an interaction between the prior outcome $Y_{0}$ and the original IV $Z$ to form a new IV. This general technique to use interaction terms  has been successfully applied in several different contexts in recent years. For example  to disentangle direct and indirect effects in a mediation analysis \cite{Small12}, to allow for violation of the homogeneity assumption in a non-adherence affected RCT \cite{Bowden21b}, and to adjust for bias due to pleiotropy in Mendelian randomization  \cite{Spiller19}. The POA-IV estimates follows the same idea as the mentioned studies using interaction terms as instruments.
\\
\\
The treatment effect, $\beta_{POA-IV}$ can be estimated using a slightly modified TSLS approach. In the first stage model, treatment assignment $X$ is regressed on $Z$ and $W_1$, but also on $Y_0$ and the interaction term $Y_0Z$ as the new IV:

\begin{equation}\label{equ:exIV_1}
    \text{Logit}(P(X=1|Z, W_1, Y_0)) = \alpha_{X,0} + \alpha_{X,Z}Z + \alpha_{X,W_1} W_1 + \alpha_{X,Y_0} Y_0 + \alpha_{X,Y_0Z}Y_0 Z.
\end{equation}

\noindent By including $Y_0$ in the first stage model, the estimate acknowledges that the treatment decision can be affected by previously measured outcomes such as drug specific adverse events or other outcomes in the prior period. Fitted values from regression model (\ref{equ:exIV_1}), $\hat{X}$, are then used in the second stage model:

\begin{equation}\label{equ:exIV_2}
    E(Y_1|\hat{X}, W_1, Y_0, Z) = \beta_{Y_1,0} + \beta_{POA-IV}\hat{X} + \beta_{Y_1,W_1}W_1 + \beta_{Y_1,Z}Z + \beta_{Y_1,Y_0}Y_{0} ,
\end{equation}

\noindent to furnish a causal estimate for $X$ whilst additionally controlling for any direct effects of $Z$ and $Y_{0}$ on $Y_{1}$. The interaction term $\alpha_{X,Y_0Z}$ in (\ref{equ:exIV_1})  would be present in our setting if a practitioner only shows a prescription preference in situations where the patient has already experienced an event of interest in the prior period or, more generally, if the strength of preference varies across levels of $Y_0$. As it serves as a new IV we require that  
\begin{itemize}
    \item{POA-IV1: The interaction term $\alpha_{X,Y_0Z}$ is non-zero and strong (in order to avoid weak instrument bias.  \cite{Small12})};
    \item{POA-IV2: $Z$ and  $Y_{0}$ (and hence $\hat{X}$ in model (\ref{equ:exIV_2})) are independent of $U$.}
\end{itemize}
To implement the approach using the CF model for binary outcomes (which we refer to as POA-CF), we again fit model (\ref{equ:exIV_1}) to the data to give $\hat{X}$. From this we calculate the residual $\hat{\Delta} = X - \hat{X}$,  and then fit the second stage regression model 
\begin{equation}\label{equ:POACF_2}
    \text{Logit}(P(Y_1=1|X, W_1, Y_0, Z, \hat{\Delta})) = \beta_{Y_1,0} + \beta_{POA-CF}X + \beta_{Y_1,W_1}W_1 + \beta_{Y_1,Z}Z + \beta_{Y_1,Y_0}Y_0 
\end{equation}
\begin{equation*}
    + (\beta_{Y_1, \hat{\Delta}} +  \beta_{Y_1, Z \hat{\Delta}}Z) \hat{\Delta},
\end{equation*}
before estimating the causal effect on the risk difference scale using the {\tt  margins()} package as before. The performance of both the POA-CF and equivalent standard IV method (referred to as POA-IV)  are explored in the next section.

\subsection{Simulation study}\label{sec:sim2}
\begin{figure}[!htbp]
	\centering
	\adjincludegraphics[height=17cm,trim={0cm 0cm 0cm 0cm},clip]{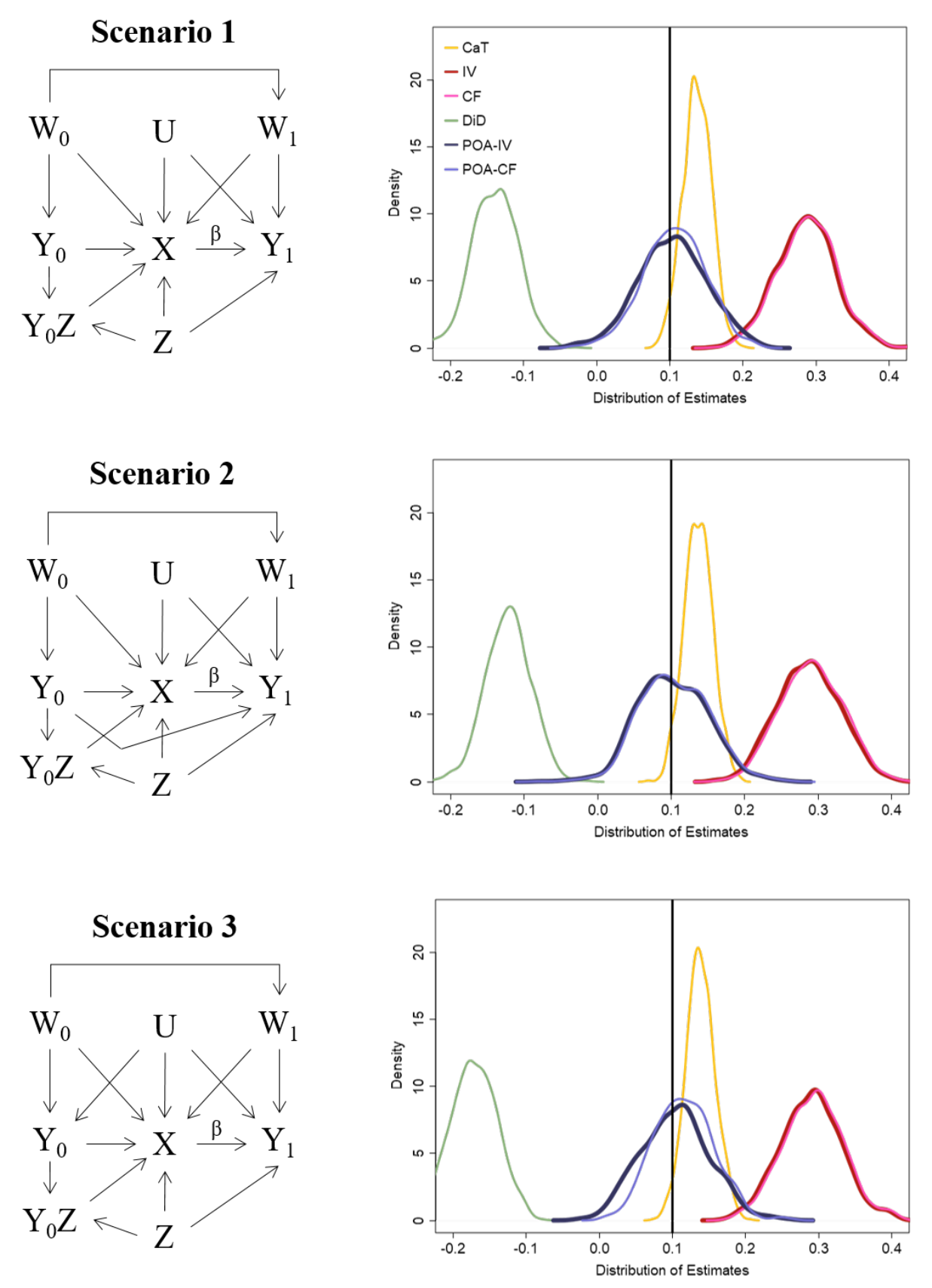}
	\caption{Left side: DAG representing the data generation for each scenario of the simulation. Right side:  Distribution of the estimation results.}
	\label{fig:scenariossimuation2}
\end{figure}

We now showcase the ability of the POA-IV and POA-CF estimate in comparison to the CaT, IV, CF and DiD estimates under conditions in which the latter three approaches are biased. The simulation is therefore an extension of the simulation described in Section \ref{sec:sim1}. The left side of Figure \ref{fig:scenariossimuation2} clarifies how the data for each scenario is generated. For scenario 1 and 2 of this simulation study the prior outcome $Y_0$ is generated without unmeasured confounding. Additionally, in scenario 2 $Y_0$ has a direct effect on the study outcome of interest $Y_1$. Scenario 3 is the same as scenario 8 of the simulation in Section \ref{sec:sim1}. 
Further information about the data generation models are outlined in Appendix 4.

\subsection{Simulation results}

The results of scenario 1 in Table \ref{tab:results_sim2} and the right side of Figure \ref{fig:scenariossimuation2} show that the POA-IV and POA-CF are able to estimate the true causal risk difference ($10\%$)  without bias and similar efficiency to the standard IV estimate. All other methods compared in this simulation exhibit bias. This is also the case for scenario 2 in which $Y_0$ exerts a direct effect on $Y_1$. For this scenario $Y_0$ was included in the outcome models as measured confounder. Coverage rates of the POA-IV and POA-CF are around $95\%$.  The bias of the IV and CF approach in all scenarios stems from a relatively large effect of $Z$ on $Y_1$. Scenario 3 of this simulation is the same as scenario 8 described in Section \ref{sec:sim1}. All previously applied methods exhibited noticeable bias. As $Y_0$ is confounded with $U$ in this scenario, POA-IV as well as  POA-CF are biased too. From the results of this simulation the bias is much smaller than the bias of CaT, IV, CF and DiD, but it would increase in case of a stronger effect of U on $Y_0$.

Additional information on the Monte Carlo simulation errors are given in Appendix 5. 

\begin{longtable}{l|lllllll}
\hline
\multicolumn{1}{c|}{  } &   & CaT     & \multicolumn{1}{c}{IV} & \multicolumn{1}{c}{CF}  & \multicolumn{1}{c}{DiD} & \multicolumn{1}{c}{POA-IV} & \multicolumn{1}{c}{POA-CF}\\ \hline
\endhead
\multirow{5}{*}{Scenario 1}    & bias & 3.72 & 18.6 & 18.9 & -23.9 & 0.369 & 0.636 \\ 
 & S.E. & 0.0615 & 0.129 & 0.129 & 0.101 & 0.15 & 0.136 \\ 
 & MSE & 0.176 & 3.62 & 3.73 & 5.82 & 0.225 & 0.188 \\ 
 & coverage & 52.8 & 0.4 & 0.4 & 0 & 94.4 & 95.2 \\ 
 & T1E & 44.4 & 99.4 & 99.4 & 100 & 5.9 & 5.9 \\   \hline
\multirow{5}{*}{Scenario 2}    & bias & 3.71 & 18.7 & 19.1 & -22.2 & 0.219 & 0.527 \\ 
  & S.E. & 0.0613 & 0.135 & 0.135 & 0.0974 & 0.15 & 0.15 \\ 
  & MSE & 0.175 & 3.69 & 3.85 & 5.02 & 0.227 & 0.228 \\ 
  & coverage & 53.6 & 0.6 & 0.5 & 0 & 95.7 & 95.6 \\ 
  & T1E & 45.7 & 99.3 & 99.4 & 100 & 4 & 4.2 \\   \hline
\multirow{5}{*}{Scenario 3}    & bias & 3.85 & 19 & 19.3 & -27.1 & 0.428 & 1.59 \\ 
 & S.E. & 0.0629 & 0.131 & 0.131 & 0.0998 & 0.149 & 0.131 \\ 
 & MSE & 0.187 & 3.77 & 3.9 & 7.45 & 0.225 & 0.196 \\ 
 & coverage & 50.7 & 0.3 & 0.3 & 0 & 95.5 & 92.8 \\ 
 & T1E & 48.9 & 99.7 & 99.7 & 100 & 6 & 6.2 \\ \\           
\caption{ Bias, standard errors (SE) and mean squared error (MSE) (all $\times$ 100);  coverage and type 1 error (T1E) rate (both expressed as a percentage based on a 95\% confidence interval and 5\% significance threshold) for the estimates CaT, IV, CF, DiD, POA-IV and POA-CF and for all scenarios. }\label{tab:results_sim2}
\end{longtable}

\section{Application to Type 2 Diabetes patients in Clinical Practice Research Datalink} \label{sec:application}
In addition to lifestyle modification, treatment for Type 2 Diabetes (T2D) primarily focuses on the management of blood glucose, with different glucose-lowering oral agents  available. Metformin (MFN) is recommended as first-line medical therapy by major T2D clinical guidelines \cite{NICE20, Buse20}, but if glucose control deteriorates, additional second-line or further treatments are prescribed. Sodium-glucose co-transporter-2 inhibitors (SGLT2i) and Dipeptidyl peptidase-4 inhibitors (DPP4i) are two widely used second-line medication classes in the UK and US \cite{Dennis19b, Montvida18}, and there is considerable interest in using observational data to establish the comparative benefits and risks of the two therapies in `real-world' settings and for a broad spectrum of patients. \cite{Dennis20} Whilst SGLT2 inhibitors have some benefits beyond blood sugar lowering (including reducing the risk of cardiovascular disease) they may be associated with increased risk for genital infection. \cite{Buse20}
\\ 
\\
We used routine data from the Clinical Practice Research Datalink (CPRD) to examine the risk of genital infections for people with T2D initiating SGLT2i (n = 1966) compared to DPP4i (n = 4033) during 2016-2019 as second-line treatment after MFN. \cite{Herrett15, Rodgers17} CPRD is a rich source of primary care data for observational health research. This database includes approximately 6.9\% of the UK population and patients are considered to be representative with regard to age, sex and ethnicity. \cite{Herrett15} Studying the efficacy and tolerability with routine practice data makes it possible to understand the risks and benefits of medication use in a large and truly representative population in contrast to clinical trials which are performed on a population restricted by factors such as age or diabetes severity. \cite{Booth14,Hinton18} 
\\
\\
All individuals in the study cohort initiated MFN as first-line treatment and had not been prescribed insulin over the complete follow-up time. Additionally, only individuals who initiated DPP4i or SGLT2i as second-line treatment were included in the analysis. The prior period is observed from start of the initiation of MFN until just before the start of the second-line treatment (SGLT2i or DPP4i). The average follow-up time in this period was 3.66 years. Therefore, the study period starts with the initiation of the second-line treatment until one of the following-censoring reasons: end of follow-up data (30th of June 2019), discontinuation of second-line treatment or start of the other comparison treatment (eg.: individual started to take SGLT2i as second-line treatment and added DPP4i at a later point in time). The average follow-up time of the study period was 1.44 years. 
\\
\\
The baseline characteristics of the cohort are summarized in Table \ref{tab:cohort_description}, for the two periods before and after initiation of second-line treatment. In the prior period 151 (2.5\%) genital infections are recorded, 45 (2.3\%) and 106 (2.6\%) for people on SGTL2i and DPP4i respectively. In the study period 139 (2.3\%)  people experience an infection, 96 (4.9\%) on SGLT2i and 43 (1.1\%) on DPP4i. Genital infection is therefore a rare outcome. Data was also extracted on patients' general practice membership, in order to use it as an IV within the standard and prior outcome augmented IV approach.\\
\\
The outcome, defined as  $\geq$ 1 genital infection in a given period, was  coded as a binary variable and modelled using logistic regression. Causal estimates are reported on the risk difference scale (in \%) as described in Section \ref{sec:binary_outcome}. 
\\
\\
For our analysis we applied the six causal estimation strategies introduced in Sections \ref{sec:methods} and \ref{sec:exIV} to estimate the population averaged effect of taking SGLT2i versus DPP4i  on infection risk. Additionally, we applied the CaT estimator on a propensity score matched dataset (PSM) using the R package {\tt  MatchIt()} with 1-1 nearest neighbour matching. \cite{Ho11} Approximately two-thirds of the data was matched. The balance diagnostic statistics are summarized with a love-plot in Appendix 6 and show that the matching procedure has improved the balance of the treatment groups. This plot also gives a list of the variables which was used for the matching procedure. Furthermore, the CaT and DiD method are applied including $Z$ as measured confounder to avoid bias in case the exclusion restriction of the IV method is not met, which cannot be verified with the data at hand. \cite{Lousdal18} $Y_0$ was included as a measured confounder in all models for $Y_1$ as it has been found in previous studies that prior infections are associated with the risk of experiencing an infection in the study period. \cite{McGovern20} The $\beta_{CaT}$ estimate was obtained from a multivariable logistic regression adjusted for all baseline characteristics measured at second-line treatment initiation, as listed in Table \ref{tab:cohort_description}. The $\beta_{DiD}$ estimate was obtained using logistic difference-in-difference regression and also adjusted for the baseline characteristics at initiation of first- and second-line treatment. Standard IV, CF and the prior outcome augmented IV approaches were fitted to the data using the methods previously described, with adjustment for the same set of baseline covariates in the first and second stage models.

\subsection{Construction of the Instrumental Variable}

As prescription prevalence of both drug classes increased dramatically after 2015 and regional differences in prescribing patterns in the UK exist \cite{Curtis18, Dennis19b}, the IV $Z$ constructed for this analysis aims to convey information about a practitioner's preference to prescribe SGTL2i over DPP4i. Preference-based IVs have been proposed when it is assumed that physicians prescription preference varies or a substantial variation in practice pattern can be observed. \cite{Korn98, Brookhart06b} Hence, for the estimation of $\beta_{IV}$, $\beta_{CF}$,  $\beta_{POA-IV}$ and $\beta_{POA-CF}$, we constructed a binary IV for each practice according to the following rule: $Z_j = 0$  for all patients of practice $j$ if the last patient treated at practice $j$ was prescribed DPP4i and $Z_j = 1$ if the last patient of practice $j$ was prescribed SGLT2i. The practitioner is assumed to have a preference to prescribe SGLT2i over DPP4i depending on the most recent prescription. The patient data used to construct this variable was then excluded from the analysis. As SGLT2i and DPP4i are newer drug classes and started to be prescribed as second-line treatment more often after 2014 \cite{Dennis19b, Montvida18}, we allowed for an initial period in which preference could develop. Therefore, data of individuals initiating second-line treatment from 2016 onward is analysed. A similar approach of using clinical commissioning group prescribing history as preference based IV has been proposed to evaluate T2D treatment by Bidulka et al. (2021). \cite{Bidulka21}

\begin{table}[h]\centering
\begin{tabular}{l|ll|ll}
\hline
\multicolumn{1}{c|}{}            & \multicolumn{2}{c|}{prior period}                       & \multicolumn{2}{c}{study period}                       \\ \hline
\multicolumn{1}{c|}{Variable}    & \multicolumn{1}{c}{DPP4i} & \multicolumn{1}{c|}{SGLT2i} & \multicolumn{1}{c}{DPP4i} & \multicolumn{1}{c}{SGLT2i} \\ \hline
HbA1c (mmol/mol)                 & 70.68 (19.0)             & 72.54 (19.69)               & 69.93 (15.24)             & 74.18 (16.22)              \\
BMI (kg/m$^2$)                   & 33.31 (6.32)              & 36.53 (6.81)                & 32.51 (6.36)              & 35.8 (6.7)               \\ 
eGFR (mL/min)                   & 84.91 (19.88)              & 92.81 (18.25)               & 83.11 (23.05)             & 93.08 (18.67)              \\
age (years)                     & 60.96 (11.0)               & 55.19 (8.96)                & 64.85 (11.6)             & 58.48 (9.18)                \\
T2D duration (years)             & 2.28 (3)                  & 1.82 (2.59)                 & 6.13 (4.34)               & 5.09 (3.6)                \\
gender                           &                           &                             &                           &                            \\
\multicolumn{1}{r|}{female}      & 41\%                  & 60.9\%                    &                           &                            \\
\multicolumn{1}{r|}{male}        & 59\%                  & 39.1\%                    &                           &                            \\
prescription year                &                           &                             &                           &                            \\
\multicolumn{1}{r|}{2016} &                           &                             & 35.53\%                   & 23.3\%                       \\
\multicolumn{1}{r|}{2017}  &                           &                             & 30.3\%                   & 32.91\%                      \\
\multicolumn{1}{r|}{2018}  &                           &                             & 27.23\%                   & 33.83\%                      \\
\multicolumn{1}{r|}{2019}  &                           &                             & 6.94\%                   & 9.97\%                      \\

\end{tabular}
\caption{\label{tab:cohort_description} {Baseline data on CPRD T2D cohort for prior and study period and for patients on DPP4i (n = 4033) or SGLT2i (n = 1966) as second-line treatment. Values are shown in mean (standard deviation) unless otherwise stated.}}
\end{table}

%----------------------------------------------------------------

\subsection{Results}

\begin{figure}[!htbp]
	\centering
	\includegraphics[width=0.6\linewidth]{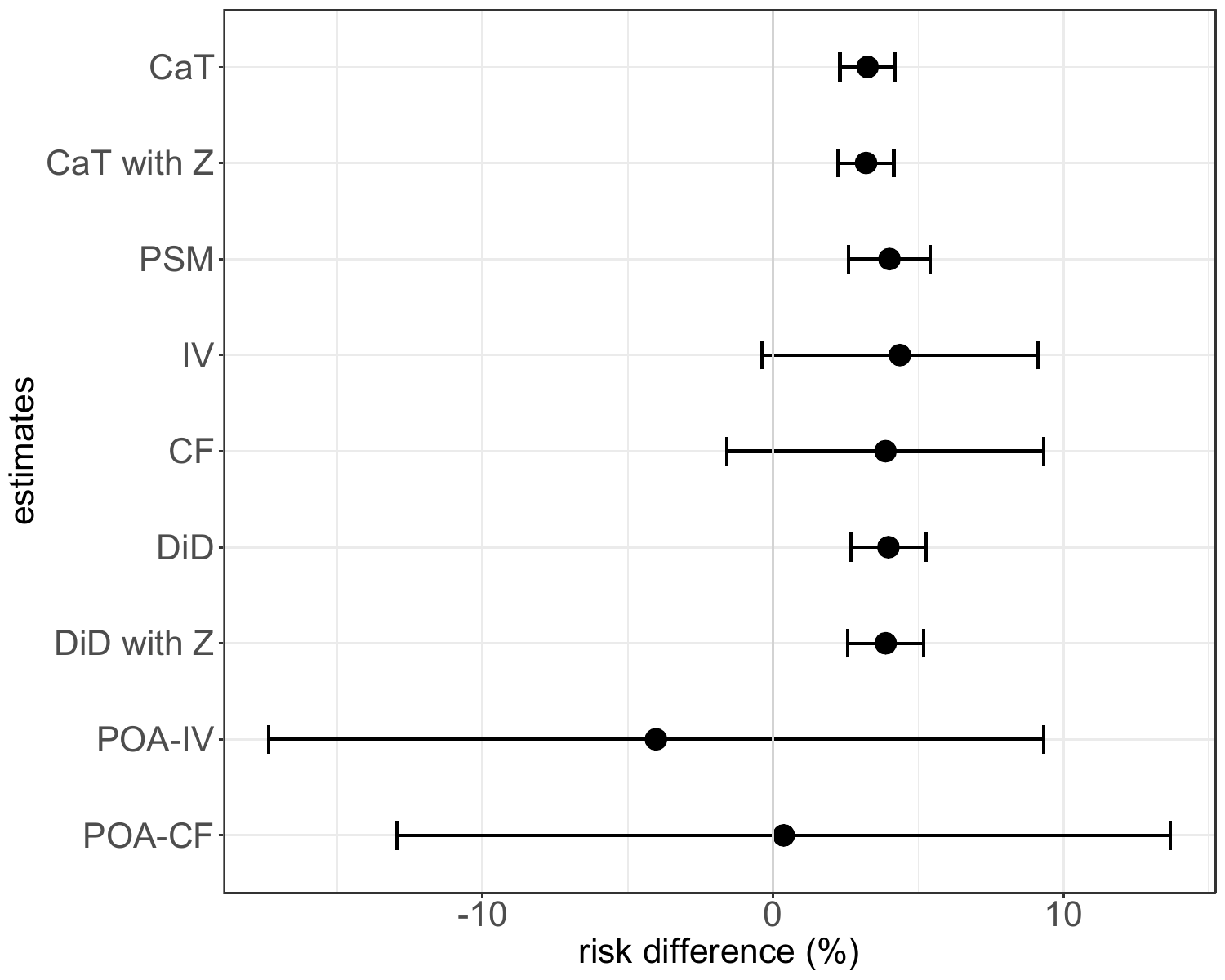}
	\caption{Estimated treatment effect for all estimates and their 95\% confidence intervals.}
	\label{fig:pointesttimates_CIs}
\end{figure}

\noindent The results of the causal analysis are given in Table \ref{tab:applied_analysis_results} and  Figure \ref{fig:pointesttimates_CIs}. 
All methods except the POA-IV approach estimate a positive causal effect suggesting that genital infection risk is higher if all people initiated SGLT2i compared to DPP4i. The standard IV, CF POA-IV and POA-CF causal estimates are not significantly different from zero at or below the 5\%  significance threshold.
The POA-IV estimates a negative causal effect, but its uncertainty is large compared to all other approaches and consequently its 95\% confidence interval crosses the null. This is also true for the POA-CF estimate. Although the POA-IV and POA-CF estimate can deal with a direct effect of the prior outcome $Y_{0}$ on future treatment $X$, it assumes non unmeasured confounding between $Y_{0}$ and $X$. Including $Z$ in the DiD and CaT model does not result in a big change of the estimation results in this application study. 

\begin{table}[h]\centering
\begin{tabular}{l|llll}
\hline
\multicolumn{1}{c|}{Method} & \multicolumn{1}{c}{Estimate} &  \multicolumn{1}{c}{95\% CI} & \multicolumn{1}{c}{S.E.} & \multicolumn{1}{c}{p-value} \\ \hline
CaT                     & 3.25            &  2.30; 4.21        & 0.49     & $2.08\times10^{-11}$  \\
CaT with Z              & 3.2            &  2.25; 4.16        & 0.49     & $5.41\times10^{-11}$   \\
PSM                     & 4.0            &  2.6; 5.42      & 0.72     & 0.04  \\
IV                      & 4.38            &  -0.35; 9.12        & 2.42     & 0.07  \\
CF                      & 3.93            &  -1.5; 9.38        & 2.77     & 0.16  \\
DiD                     & 3.97            &  2.67; 5.27      & 0.66     & $2.21\times10^{-9}$ \\
DiD with Z              & 3.88            &  2.57; 5.19      & 0.67     &  $7.1\times10^{-9}$ \\
POA-IV                    & -4.03           &  -17.38; 9.31    & 6.81     & 0.55  \\    
POA-CF                    & 0.37           &  -12.94; 13.69    & 6.79     & 0.96
\end{tabular}
\caption{\label{tab:applied_analysis_results} {Estimation results on risk difference scale (in $\%$), standard error, and p-value of the estimated treatment effect.}}
\end{table}

\noindent Table \ref{tab:IV strength} summarized the strength of the IVs measured with the F-statistic of coefficient of each IV from the first stage regressions of each respective method. \cite{Stock02} IV and CF as well as POA-IV and POA-CF use the same first stage regression model and the results of the IV strength are therefore summarized in the same row.  The instrument strength of $Z$ for the IV and CF approach is strong with F-staitsitcs greater than $10$ but $Y_0Z$ does not seem to be a strong instrument for the POA-IV and POA-CF approach. This helps to understand the poor results of the two methods which are not in line with the estimation results of all other methods applied in this study. Furthermore, it is plausible that $Y_0$ is confounded by $U$ which will also lead to biased results for the POA-IV and POA-CF.

\begin{table}[h]\centering
\begin{tabular}{l|ll}
\hline
\multicolumn{1}{c|}{Models} & \multicolumn{1}{c}{instrument} & \multicolumn{1}{c}{F-statistics} \\ \hline
IV and CF & $Z$         & 144.18    \\
POA-IV and POA-CF  & $Y_0Z$        & 1.61    \\
\end{tabular}
\caption{\label{tab:IV strength} {Strength of the instrumental variables measured with the F-statistics of $Z$ (for IV and CF) and $Y_0Z$ (POA-IV and POA-CF) from the corresponding first stage regression models.}}
\end{table}

\subsection{Results of the similarity statistics}  
We now apply our $Q_e$ statistic analysis to the set of estimators to assess their similarity. 
Table \ref{tab:Q_statistics_results} shows the $Q_{e}$ statistic for a selection of estimator  sets. The pairwise correlation of all estimates calculated over 500 bootstrap samples is summarized in Figure \ref{fig:correlation_plot} in Appendix 7. Interestingly, the test statistic for the closely related CaT and PSM estimates as well as for the POA-IV and POA-CF estimates reveal they are not sufficiently similar even though their values are very close. This is explained by their very high correlation, which $Q_{e}$ adjusts for.  IV and CF are identified as sufficiently similar even after accounting for their correlation. All other selected combinations which do not include CaT or PSM and POA-IV or POA-CF together show that the estimates are statistically similar.
\begin{table}[h]\centering
\begin{tabular}{l|llll}
\hline
\multicolumn{1}{c|}{${e}$} & \multicolumn{1}{c}{$Q_{e}$ statistic} & \multicolumn{1}{c}{$\chi^2$ value (df)} & \multicolumn{1}{c}{p value} & \multicolumn{1}{c}{test decision}       \\ \hline
CaT, PSM          & 9.721    &  3.841 (1) &  0.0018  & not similar \\
IV, CF          & 0.093    &  3.841 (1) &  0.761 & similar \\
POA-IV, POA-CF        & 9.086     & 3.841 (1) &  0.0026 & not similar \\
CaT, CF, DiD, POA-CF     & 2.848     & 7.815 (3) &  0.4157 & similar \\
PSM, CF, DiD, POA-CF        & 0.343     & 7.815 (3) &  0.9517  & similar \\
CaT, IV, DiD, POA-IV        & 4.101     & 7.815 (3) &  0.2508  & similar \\
PSM, IV, DiD, POA-IV       & 1.367     & 7.815 (3) &  0.7132  & similar 
\end{tabular}
\caption{\label{tab:Q_statistics_results} {Test results of the heterogeneity test with $Q_{e}$ statistic and 95\% confidence.}}
\end{table}

\section{Summary and conclusion}\label{sec:discussion}

% summary:
In this paper we propose a framework for the application of several causal inference methods to assess the comparative effectiveness of two treatments in observational data. This included `standard' confounder adjustment approaches such as multivariable regression and propensity score matching, Difference-in-Differences and IV estimation. The assumptions of each approach were  described, and a simulation was used to assess the impact of violating necessary assumptions on the estimators' performance. Building on the work of Bowden et. al. \cite{Bowden21a}, we proposed the use of a similarity statistic to formally assess the level of agreement between sets of estimates that can account for their underlying correlation. We hope this statistic could be useful tool when attempting to triangulate findings from a set of distinct causal estimation strategies going forward.
\\
\\
We illustrated the application of these methods using routinely collected data on people with T2D, to assess the relative benefit of SGLT2i compared to DPP4i as second-line therapies on the risk of genital infection. Our heterogeneity analysis showed good agreement between all causal estimates except the PSM/ CaT and POA-IV/POA-CF approaches. In future work, we plan to apply the same causal framework to model alternative T2D outcomes such as HbA1c and other clinically important adverse events. We also plan to extend the approach to fit alternative models that allow for causal effect heterogeneity, so that they may be used in personalised medicine.\cite{Dennis21} Furthermore, the applied analysis showcased how triangulating estimation results with different methods can help to identify  implausible results such as the results of POA-IV/ POA-CF  and in this application give further insight in possible assumption violations. Additional work still needs to be done about which estimation results final reports should focus on. Possible strategies could be to only focus on similar estimates or to combine the results. 
\\ 
\\
We proposed the use of the POA-IV/ POA-CF method which is able to leverage an interaction between the prior period and the IV accounting for a possible direct effect of the IV on the outcome and a direct effect of previous outcome events on the treatment decision.  Our simulations show that this approach is robust and leads to reliable results in scenarios in which key assumptions of the DiD and the IV approaches are violated, as long as the prior outcome-future treatment relationship does not suffer from unmeasured confounding. Furthermore, our simulations in Section \ref{sec:sim2} showed that POA-IV and POA-CF were less biased than CaT, DiD, IV and CF even if $Y_0$ was confounded by $U$. As future work we hope to better understand when this will be the case. 
\\
\\
As further research, we plan to develop a rigorous hierarchical testing procedure for performing a similarity analysis across an arbitrary number of estimates, whilst controlling the overall family wise error rate. Another approach for combining IV and DiD approaches has recently been  proposed by Ye et al. 2020. \cite{Ye20} The `instrumented DiD' purports to offer robustness to time-varying unmeasured confounding and therefore offers considerable utility as an additional estimator within a triangulation analysis.

\subsection{Acknowledgements} 
This article is based in part on data from the Clinical Practice Research Datalink obtained under licence from the UK Medicines and Healthcare products Regulatory Agency. CPRD data is provided by patients and collected by the NHS as part of their care and support.

\subsection{Funding}
Approval for the study was granted by the CPRD Independent Scientific Advisory Committee (ISAC 13\_177R). LG, JMD and JB  are both supported by a grant from Research England’s Expanding Excellence in England (E3) fund.

\subsection{Conflict of interest}

The authors declare no conflict of interest.

\subsection{Author contribution}

LG and JB conceived and developed the methodological framework, with the constant supervisory support of BMS, LRR and JMD. APM provided invaluable clinical insight for the applied analyses in Section 5 and WH provided useful insight on the assumptions required by DiD regression. LG drafted the original version of the paper which all authors helped to edit. All authors read and approved the final version of the manuscript.

\subsection{Data availability statement}
Data from CPRD is available to all researchers following successful application to the ISAC. Source code for this research for all simulations and the application study in this paper is available at \url{https://github.com/GuedemannLaura/POA-IV}
\newpage
\subsection{Bibliography}
%\nocite{*}% Show all bib entries 
\bibliographystyle{SageV}
\bibliography{References}

\begin{thebibliography}{10}
\providecommand{\url}[1]{\texttt{#1}}
\providecommand{\urlprefix}{URL }
\expandafter\ifx\csname urlstyle\endcsname\relax
  \providecommand{\doi}[1]{DOI:\discretionary{}{}{}#1}\else
  \providecommand{\doi}{DOI:\discretionary{}{}{}\begingroup
  \urlstyle{rm}\Url}\fi
\providecommand{\eprint}[2][]{\url{#2}}

\bibitem{Greenland01}
Greenland S and Morgenstern H.
\newblock Confounding in health research.
\newblock \emph{Annual Review of Public Health} 2001; 22: 189--212.

\bibitem{Jager08}
Jager KJ, Zoccali C, MacLeod A et~al.
\newblock Confounding: What it is and how to deal with it.
\newblock \emph{Kidney International} 2008; 73: 256--260.
\newblock \doi{10.1038/sj.ki.5002650}.

\bibitem{Boyko13}
Boyko EJ.
\newblock Observational research - opportunities and limitations.
\newblock \emph{Journal of Diabetes and Its Complications} 2013; 27: 642--648.

\bibitem{Dawwas19}
Dawwas GK, Smith SM and Park H.
\newblock Cardiovascular outcomes of sodium glucose cotransporter‐2
  inhibitors in patients with type 2 diabetes.
\newblock \emph{Metabolism} 2019; 21: 28--36.

\bibitem{McGovern20}
McGovern AP, Hogg M, Shields BM et~al.
\newblock Risk factors for genital infections in people initiating sglt2
  inhibitors and their impact on discontinuation.
\newblock \emph{BMJ Open Diabetes Research and Care} 2020; 8.
\newblock \doi{10.1136/bmjdrc-2020-001238}.

\bibitem{Streeter17}
Streeter AJ, Lin NX, Crathorne L et~al.
\newblock Adjusting for unmeasured confounding in nonrandomized longitudinal
  studies: a methodological review.
\newblock \emph{Journal of Clinical Epidemiology} 2017; 87: 23--34.

\bibitem{Bowden21b}
Bowden J, Bornkamp B, Glimm E et~al.
\newblock Connecting instrumental variable methods for causal inference to the
  estimand framework.
\newblock \emph{Statistics in Medicine} 2021; 40: 5605--5627.

\bibitem{Penzzin15}
Pezzin LE, Laud P, Yen T et~al.
\newblock Re-examining the relationship of breast cancer hospital and surgical
  volume to mortality: an instrumental variable analysis.
\newblock \emph{Medical care} 2015; 53(12): 1033.

\bibitem{Smith03}
Smith GD and Ebrahim S.
\newblock ‘mendelian randomization’: can genetic epidemiology contribute to
  understanding environmental determinants of disease?
\newblock \emph{International Journal of Epidemiology} 2003; 32: 1--22.

\bibitem{Brookhart06b}
Brookhart MA and Wang P.
\newblock Evaluating short-term drug effects using a physician-specific
  prescribing preference as an instrumental variable.
\newblock \emph{Epidemiology} 2006; 17: 268--275.

\bibitem{Rodgers20}
Rodgers LR, Dennis JM, Shields BM et~al.
\newblock Prior event rate ratio adjustment produced estimates consistent with
  randomized trial: a diabetes case study.
\newblock \emph{Journal of Clinical Epidemiology} 2020; 122: 78--86.
\newblock \doi{10.1016/j.jclinepi.2020.03.007}.

\bibitem{Lin16}
Lin NX and Henley WE.
\newblock Prior event rate ratio adjustment for hidden confounding in
  observational studies of treatment effectiveness: a pairwise cox likelihood
  approach.
\newblock \emph{Statistics in Medicine} 2016; 35: 5149--5169.
\newblock \doi{10.1002/sim.7051}.

\bibitem{Stock02}
Stock J, Wright JH and Yogo M.
\newblock A survey of weak instruments and weak identification in generalized
  method of moments.
\newblock \emph{Journal of Business \& Economic Statistics} 2002; 20: 518--529.

\bibitem{Aso20}
Aso S and Yasunaga H.
\newblock Introduction to instrumental variable analysis.
\newblock \emph{Annals of Clinical Epidemiology} 2020; 2: 69--74.

\bibitem{Lousdal18}
Lousdal ML.
\newblock An introduction to instrumental variable assumptions, validation and
  estimation.
\newblock \emph{Emerging Themes in Epidemiology} 2018; 15: 1--7.
\newblock \doi{10.1186/s12982-018-0069-7}.

\bibitem{Wooldridge19}
Wooldridge JM.
\newblock \emph{Introductory Econometrics - A Modern Approach}.
\newblock 7th ed. South-Western College Publishing, 2019.

\bibitem{Bowden84}
Bowden RJ and Turkington DA.
\newblock \emph{Instrumental Variables}.
\newblock Cambridge: Cambridge University Press, 1984.

\bibitem{Weiner08}
Weiner MG, Xie D and Tannen RL.
\newblock Replication of the scandinavian simvastatin survival study using a
  primary care medical record database prompted exploration of a new method to
  address unmeasured confounding.
\newblock \emph{Pharmacoepidemiology and drug safety} 2008; 17: 661--670.

\bibitem{Tannen08}
Tannen RL, Weiner MG and Xie D.
\newblock Replicated studies of two randomized trials of
  angiotensin‐converting enzyme inhibitors: further empiric validation of the
  ‘prior event rate ratio’to adjust for unmeasured confounding by
  indication.
\newblock \emph{Pharmacoepidemiology and Drug Safety} 2008; 17: 671--685.

\bibitem{Uddin15}
Uddin MJ, Groenwold RHH, Staa TPV et~al.
\newblock Performance of prior event rate ratio adjustment method in
  pharmacoepidemiology: a simulation study.
\newblock \emph{Pharmacoepidemiology and drug safety} 2015; 24: 468--477.

\bibitem{Gallagher09}
Gallagher A, de~Vries F and van Staa T.
\newblock Prior event rate ratio adjustment: a magic bullet or more of the
  same?
\newblock JOHN WILEY \& SONS LTD THE ATRIUM, SOUTHERN GATE, CHICHESTER PO19
  8SQ, W ….
\newblock ISBN 1053-8569, pp. S14--S15.

\bibitem{Yu12}
Yu M, Xie D, Wang X et~al.
\newblock Prior event rate ratio adjustment: numerical studies of a statistical
  method to address unrecognized confounding in observational studies.
\newblock \emph{pharmacoepidemiology and drug safety} 2012; 21: 60--68.

\bibitem{Zhou16}
Zhou H, Taber C, Arcona S et~al.
\newblock Difference-in-differences method in comparative effectiveness
  research: Utility with unbalanced groups.
\newblock \emph{Applied Health Economics and Health Policy} 2016; 14: 419--429.

\bibitem{Bernal19}
Bernal JL, Cummins S and Gasparrini A.
\newblock Difference in difference, controlled interrupted time series and
  synthetic controls.
\newblock \emph{International journal of epidemiology} 2019; 48: 2062--2063.

\bibitem{Craig12}
Craig P, Cooper C, Gunnell D et~al.
\newblock Using natural experiments to evaluate population health
  interventions: new medical research council guidance.
\newblock \emph{J Epidemiol Community Health} 2012; 66: 1182--1186.

\bibitem{Rockers15}
Rockers PC, Røttingen JA, Shemilt I et~al.
\newblock Inclusion of quasi-experimental studies in systematic reviews of
  health systems research.
\newblock \emph{Health Policy} 2015; 119: 511--521.

\bibitem{Soumerai15}
Soumerai SB, Starr D and Majumdar SR.
\newblock How do you know which health care effectiveness research you can
  trust? a guide to study design for the perplexed.
\newblock \emph{Preventing chronic disease} 2015; 12.

\bibitem{Techetgen14}
Tchetgen ET.
\newblock A note on the control function approach with an instrumental variable
  and a binary outcome.
\newblock \emph{Epidemiologic methods} 2014; 3: 107--112.

\bibitem{Leeper21}
Leeper TJ.
\newblock \emph{margins: Marginal Effects for Model Objects}, 2021.
\newblock R package version 0.3.26.

\bibitem{Huitfeldt19}
Huitfeldt A, Stensrud MJ and Suzuki E.
\newblock On the collapsibility of measures of effect in the counterfactual
  causal framework.
\newblock \emph{Emerging Themes in Epidemiology} 2019; 16: 1--5.

\bibitem{Slotani64}
Slotani M.
\newblock Tolerance regions for a multivariate normal population.
\newblock \emph{Annals of the Institute of Statistical Mathematics} 1964; 16:
  135--153.

\bibitem{Cochran54}
Cochran WG.
\newblock The combination of estimates from different experiments.
\newblock \emph{Biometrics} 1954; 10: 101--129.

\bibitem{Bowden21a}
Bowden J, Pilling LC, Türkmen D et~al.
\newblock The triangulation within a study (twist) framework for causal
  inference within pharmacogenetic research.
\newblock \emph{PLoS Genetics} 2021; 17: e1009783.

\bibitem{Morris19}
Morris TP, White IR and Crowther MJ.
\newblock Using simulation studies to evaluate statistical methods.
\newblock \emph{Statistics in Medicine} 2019; 38: 2074--2102.
\newblock \doi{10.1002/sim.8086}.

\bibitem{Small12}
Small D.
\newblock Mediation analysis without sequential ignorability: Using baseline
  covariates interacted with random assignment as instrumental variables.
\newblock \emph{Journal of Statistical Research} 2012; 46: 91--103.

\bibitem{Spiller19}
Spiller W, Slichter W, Bowden J et~al.
\newblock Detecting and correcting for bias in mendelian randomization analyses
  using gene-by-environment interactions.
\newblock \emph{International Journal of Epidemiology} 2019; 48: 702--712.

\bibitem{NICE20}
{National Institute for Health and Care Excellence}.
\newblock Type 2 diabetes in adults: management, 2020.
\newblock \urlprefix\url{www.nice.org.uk/guidance/ng28}.

\bibitem{Buse20}
Buse JB, Wexler DJ, Tsapas A et~al.
\newblock 2019 update to: Management of hyperglycemia in type 2 diabetes, 2018.
  a consensus report by the american diabetes association (ada) and the
  european association for the study of diabetes (easd).
\newblock \emph{Diabetes Care} 2020; 43: 487--493.

\bibitem{Dennis19b}
Dennis JM, Henley WE, McGovern AP et~al.
\newblock Time trends in prescribing of type 2 diabetes drugs, glycaemic
  response and risk factors: A retrospective analysis of primary care data.
\newblock \emph{Diabetes, Obesity and Metabolism} 2019; 21: 1576--1584.

\bibitem{Montvida18}
Montvida O, Shaw J, John JA et~al.
\newblock Long-term trends in antidiabetes drug usage in the us: Real-world
  evidence in patients newly diagnosed with type 2 diabetes.
\newblock \emph{Diabetes Care} 2018; 41: 69--78.

\bibitem{Dennis20}
Dennis JM.
\newblock Precision medicine in type 2 diabetes: using individualized
  prediction models to optimize selection of treatment.
\newblock \emph{Diabetes} 2020; 69(10): 2075--2085.

\bibitem{Herrett15}
Herrett E, Gallagher AM, Bhaskaran K et~al.
\newblock Data resource profile: Clinical practice research datalink (cprd).
\newblock \emph{International Journal of Epidemiology} 2015; 44: 827--836.

\bibitem{Rodgers17}
Rodgers LR, Weedon MN, Henley WE et~al.
\newblock Cohort profile for the mastermind study: using the clinical practice
  research datalink (cprd) to investigate stratification of response to
  treatment in patients with type 2 diabetes.
\newblock \emph{BMJ open} 2017; 7: e017989.

\bibitem{Booth14}
Booth CM and Tannock IF.
\newblock Randomised controlled trials and population-based observational
  research: partners in the evolution of medical evidence.
\newblock \emph{British Journal of Cancer} 2014; 110: 551--555.

\bibitem{Hinton18}
Hinton W, Feher M, Munro N et~al.
\newblock Real-world prevalence of the inclusion criteria for the leader trial:
  Data from a national general practice network.
\newblock \emph{Diabetes, Obesity and Metabolism} 2019; 21: 1661--1667.
\newblock \doi{10.1111/dom.13710}.

\bibitem{Ho11}
Ho DE, Imai K, King G et~al.
\newblock {MatchIt}: Nonparametric preprocessing for parametric causal
  inference.
\newblock \emph{Journal of Statistical Software} 2011; 42(8): 1--28.
\newblock \doi{10.18637/jss.v042.i08}.

\bibitem{Curtis18}
Curtis HJ, Dennis JM, Shields BM et~al.
\newblock Time trends and geographical variation in prescribing of drugs for
  diabetes in england from 1998 to 2017.
\newblock \emph{Diabetes, Obesity and Metabolism} 2018; 20: 2159--2168.
\newblock \doi{10.1111/dom.13346}.

\bibitem{Korn98}
Korn EL and Baumrind S.
\newblock Clinician preferences and the estimation of causal treatment
  differences.
\newblock \emph{Statistical Science} 1998; 13: 209--235.

\bibitem{Bidulka21}
Bidulka P, O'Neill S, Basu A et~al.
\newblock Protocol for an observational cohort study investigating personalised
  medicine for intensification of treatment in people with type 2 diabetes
  mellitus: The permit study.
\newblock \emph{BMJ Open} 2021; 11.
\newblock \doi{10.1136/bmjopen-2020-046912}.

\bibitem{Dennis21}
Dennis JM, Young KG, McGovern AP et~al.
\newblock Derivation and validation of a type 2 diabetes treatment selection
  algorithm for sglt2-inhibitor and dpp4-inhibitor therapies based on
  glucose-lowering efficacy: cohort study using trial and routine clinical
  data.
\newblock \emph{medRxiv} 2021; .

\bibitem{Ye20}
Ye T, Ertefaie A, Flory J et~al.
\newblock Instrumented difference-in-differences.
\newblock \emph{arXiv:201103593 [statME]} ;
  \urlprefix\url{https://doi.org/10.48550/arXiv.2011.03593}.

\end{thebibliography}

\newpage
%
% Appendix ----
%
 \section*{Appendix 1}\label{sec:appendix1}
\begin{figure}[!htbp]
	\centering
	\adjincludegraphics[height=7.7cm,trim={0cm 0cm 0cm 0cm},clip]{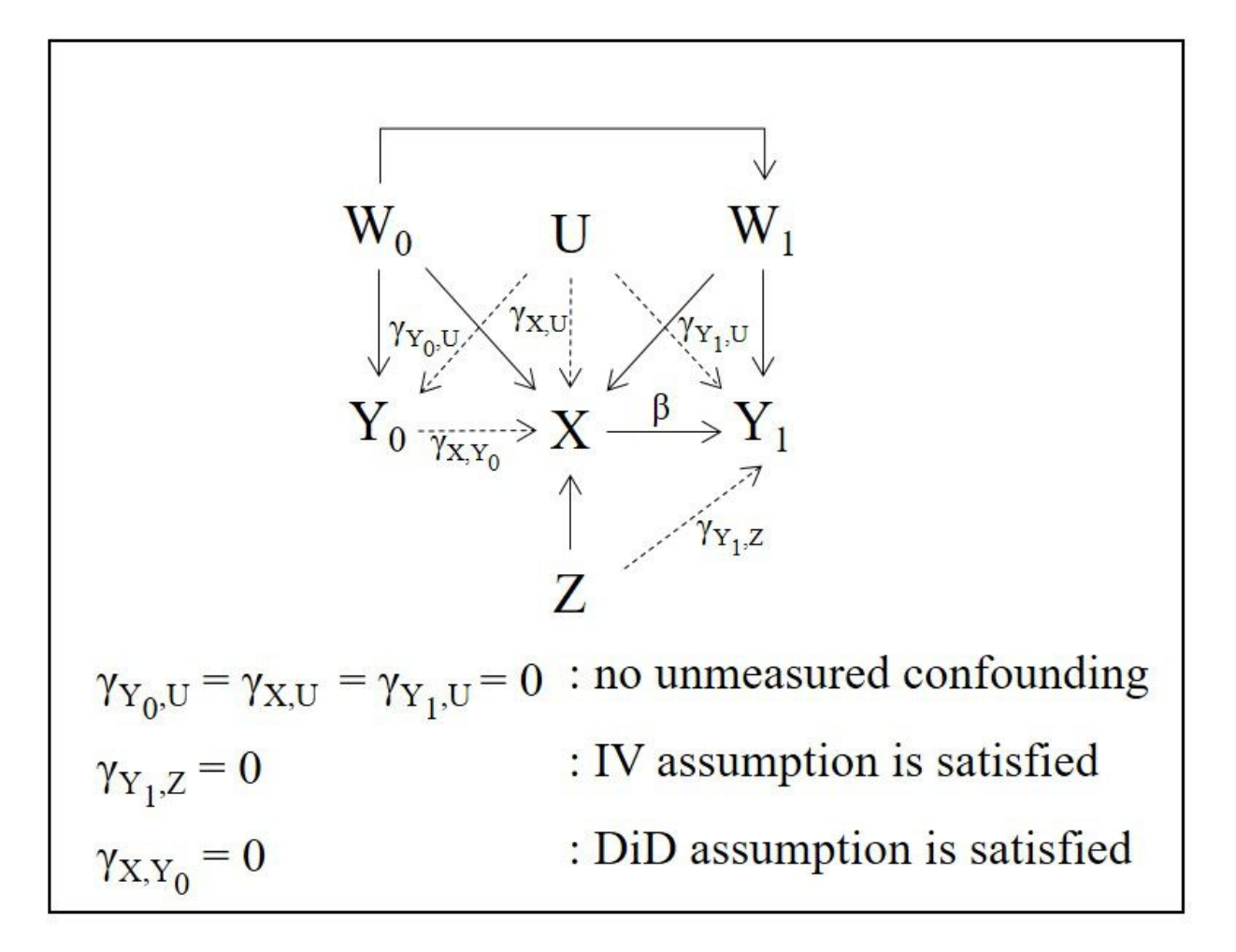}
	\caption{Causal DAG consistent with the data generation of the simulation outlined in Section \ref{sec:sim1}.}
	\label{fig:population_assumptions}
\end{figure}

Data for the first simulation described in Section \ref{sec:sim1} was generated under the models listed below. The DAG in Figure \ref{fig:population_assumptions}  visualizes the data structure of  the simulation explained in Section \ref{sec:sim1} as well as the mechanisms with which the simulation scenarios are implemented. 
\begin{table}[h]
\centering
\begin{tabular}{rcl}
 $\beta$& = & 0.1    \\
$Z_{ij}$     & $\sim$ & $Bern(0.5)$ \\ 
$W_{0,i}$ & $\sim$ & $N(0,1)$ \\ 
$W_{1,i}$ & $=$ & $\gamma_{W_1,W_0} W_{0,i} + \gamma_{W_1,\varepsilon}\varepsilon_{W_1,i} $ \\ 
$\varepsilon_{W_1,i}$ & $\sim$ & $N(0,1)$ \\
$U_i$ & $\sim$ & $N(0,1)$ \\ 
$Y_{0,i}$ & $\sim$ & $Bern(\gamma_{Y_0,0} + \gamma_{Y_0, U} U_i + \gamma_{Y_0, W_0}W_{0,i}$) \\ 
$X_i$     & $\sim$ & $Bern(\gamma_{X,0} + \gamma_{X,Z}Z_{ij} + \gamma_{X,U}U_i + \gamma_{X,W_0}W_{0,i} + \gamma_{X,W_1}W_{1,i} + \gamma_{X,Y_0} Y_{0,i}$) \\
$Y_{1,i}$ & $\sim$ & $Bern(\gamma_{Y_1,0} + \gamma_{Y_1,U}U_i + \beta X_i + \gamma_{Y_1,W_1}W_{1,i} + \gamma_{Y_1,Z}Z_{ij})$ \\ 
\end{tabular}
\end{table} 

\newpage
\section*{Appendix 2}\label{sec:appendix2}
\subsection*{Proof for DiD assumption}

\begin{figure}[!htbp]
	\centering
	\adjincludegraphics[height=4.2cm,trim={0cm 0cm 0cm 0cm},clip]{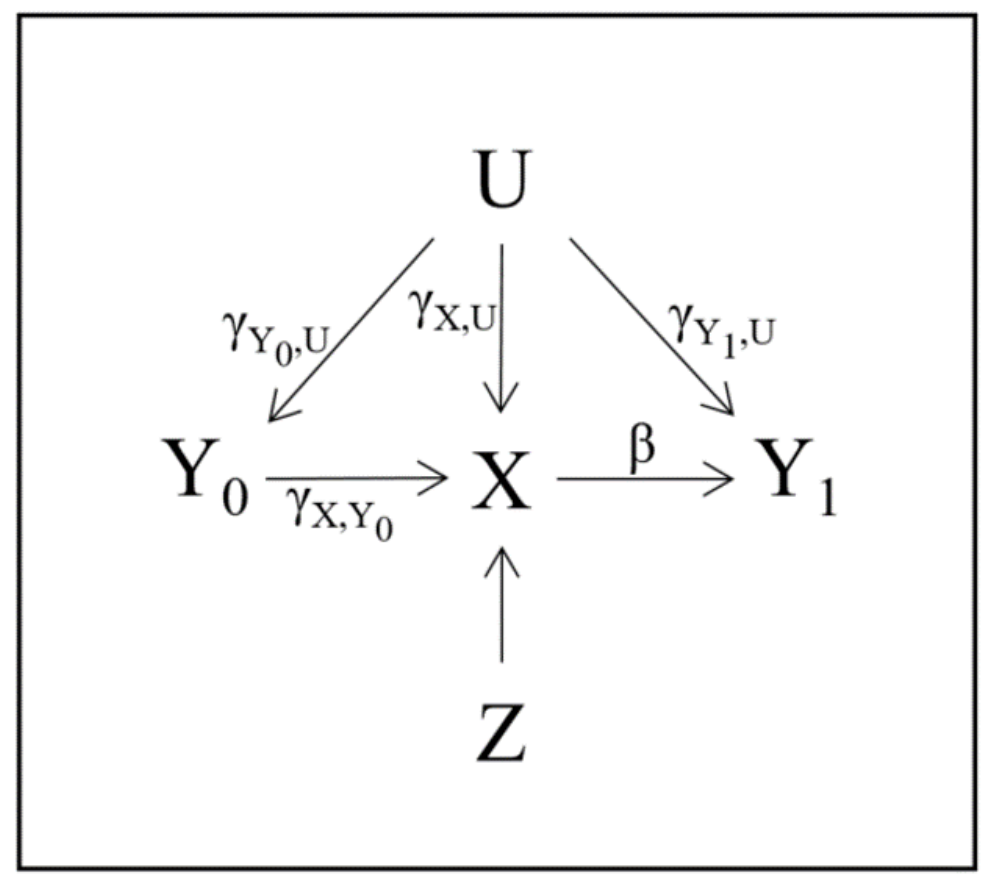}
	%\vspace{-11cm}
	\caption{{A simplified parameterised causal diagram to accompany the DiD proof argument below.}}
	\label{fig:DiD_assumption_DAG}
\end{figure} 

\noindent The parameterised causal diagram in Figure \ref{fig:DiD_assumption_DAG} indicates a similar structure to that in Section \ref{sec:introduction} but without measured confounders $W_0$ and $W_1$ for simplification. Removing the individual subscript $i$ for convenience, assume the following models for $Y_{0}$, $X$ and $Y_{1}$:

\begin{eqnarray}
Y_0 &=& \gamma_{Y_0,U}U + \epsilon_{Y0} \label{eq:Y0} \\
X &=& \gamma_{X,U}U + \gamma_{X,Y_0}Y_0+\epsilon_{X} \label{eq:X} \\
Y_1 &=& \beta X + \gamma_{Y_1,U} U + \epsilon_{Y1}, \label{eq:Y1}
\end{eqnarray}

\noindent where $\beta$ represents the causal effect that DiD is attempting to estimate. The estimand targeted by a regression of $Y_{1}$ on $X$ is therefore

\begin{equation}
\frac{Cov(Y_{1},X)}{Var(X)} = \frac{\beta Var(X)+ \gamma_{Y1,U} Cov(X,U)}{Var(X)}  \label{eq:DiD1},
\end{equation}

 \noindent and the estimand targeted by a regression of $Y_{0}$ on $X$ is therefore

\begin{equation}
\frac{Cov(Y_{0},X)}{Var(X)} = \frac{Cov(\gamma_{Y_0,U}U + \epsilon_{Y0},X)}{Var(X)}  \label{eq:DiD2}.
\end{equation}

 \noindent Putting (\ref{eq:DiD1}) and (\ref{eq:DiD2}) together, DiD estimand can be written as

\begin{equation}
\frac{Cov(Y_{1},X)}{Var(X)} - \frac{Cov(Y_{0},X)}{Var(X)} = \beta + (\gamma_{Y_{1},U}-\gamma_{Y_{0},U})\frac{Cov(U,X)}{Var(X)} - \gamma_{X,Y_{0}}\frac{Var(\epsilon_{Y_{0}})}{Var(X)}. \label{eq:DiD}
\end{equation}

\noindent From (\ref{eq:DiD}) we see that that the DiD estimand is equal to $\beta$ when $\gamma_{Y_0,U} = \gamma_{Y_1,U}$ (DiD2 assumption) and either $\gamma_{X,Y_0}$ is zero (DiD1 assumption), or that  $Var(\epsilon_{Y_{0}})$ = 0.

\newpage 

\section*{Appendix 3}\label{sec:appendix3}
For the simulation of Section \ref{sec:sim1} the Monte Carlo standard errors (MCSE) calculated based on Morris et al. 2019 \cite{Morris19}. The results are given in the table below for the performance measures: bias, mean squared error, coverage and type 1 error. 

\begin{longtable}{l|lllll}
\hline
\multicolumn{1}{c|}{  } &   & CaT     & \multicolumn{1}{c}{IV} & \multicolumn{1}{c}{CF} & \multicolumn{1}{c}{DiD} \\ \hline
\endhead
\multirow{4}{*}{Scenario 1}    & MCSE(bias) & 0.0685 & 0.1207 & 0.1209 & 0.0988 \\ 
 & MCSE(MSE) & 0.002 & 0.0064 & 0.0064 & 0.0041 \\ 
 & MCSE(coverage) & 0.7451 & 0.676 & 0.6892 & 0.7851 \\ 
 & MCSE(T1E) & 0.751 & 0.7332 & 0.7209 & 0.6892 \\    \hline
\multirow{4}{*}{Scenario 2}    & MCSE(bias) & 0.0697 & 0.1251 & 0.1252 & 0.0956 \\ 
  & MCSE(MSE) & 0.002 & 0.0071 & 0.0072 & 0.1291 \\ 
  & MCSE(coverage) & 0.6892 & 0.676 & 0.676 & 0 \\ 
  & MCSE(T1E) & 0.7085 & 0.676 & 0.676 & 0 \\    \hline
\multirow{4}{*}{Scenario 3}    & MCSE(bias) & 0.0687 & 0.1208 & 0.1209 & 0.0958 \\ 
  & MCSE(MSE) & 0.0021 & 0.0522 & 0.0529 & 0.004 \\ 
  & MCSE(coverage) & 0.7392 & 0.8628 & 0.8579 & 0.7332 \\ 
  & MCSE(T1E) & 0.6624 & 0.6957 & 0.7021 & 0.7683 \\    \hline
\multirow{4}{*}{Scenario 4}    & MCSE(bias) & 0.0685 & 0.1346 & 0.1345 & 0.1021 \\ 
 & MCSE(MSE) & 0.0022 & 0.0507 & 0.0518 & 0.148 \\ 
 & MCSE(coverage) & 0.5891 & 1.094 & 1.06 & 0 \\ 
 & MCSE(T1E) & 0.6693 & 1.0164 & 1.0126 & 0 \\    \hline
\multirow{4}{*}{Scenario 5}    & MCSE(bias) & 0.0625 & 0.1336 & 0.1332 & 0.0868 \\ 
  & MCSE(MSE) & 0.0047 & 0.0078 & 0.0078 & 0.0033 \\ 
  & MCSE(coverage) & 1.5775 & 0.676 & 0.7085 & 0.6892 \\ 
  & MCSE(T1E) & 1.5744 & 0.7451 & 0.7451 & 0.676 \\     \hline
\multirow{4}{*}{Scenario 6}    & MCSE(bias) & 0.0649 & 0.1348 & 0.1349 & 0.0935 \\ 
  & MCSE(MSE) & 0.005 & 0.0078 & 0.0078 & 0.1045 \\ 
  & MCSE(coverage) & 1.5719 & 0.5891 & 0.6197 & 0 \\ 
  & MCSE(T1E) & 1.5387 & 0.7271 & 0.7271 & 0 \\   \hline
\multirow{4}{*}{Scenario 7}    & MCSE(bias) & 0.0631 & 0.1397 & 0.1399 & 0.0894 \\ 
 & MCSE(MSE) & 0.0049 & 0.0805 & 0.0835 & 0.0036 \\ 
 & MCSE(coverage) & 1.5715 & 0.7626 & 0.7021 & 0.7147 \\ 
 & MCSE(T1E) & 1.5741 & 0.5812 & 0.5566 & 0.6957 \\   \hline
\multirow{4}{*}{Scenario 8}    & MCSE(bias) & 0.0682 & 0.1412 & 0.141 & 0.0958 \\ 
  & MCSE(MSE) & 0.0046 & 0.0768 & 0.0803 & 0.1205 \\ 
  & MCSE(coverage) & 1.5466 & 0.8278 & 0.7796 & 0 \\ 
  & MCSE(T1E) & 1.5452 & 0.7021 & 0.6826 & 0 \\     
\caption{Monte Carlo standard errors (MCSE) of the performance measures of all estimates and all scenarios of the simulation outlined in Section \ref{sec:sim1}. All results are multiplied with 100 and rounded to 3 significant figures. }\label{tab:MCSE_results_sim1}
\end{longtable}

\newpage
\section*{Appendix 4}\label{sec:appendix4}
\begin{figure}[!htbp]
	\centering
	\includegraphics[width=0.47\linewidth]{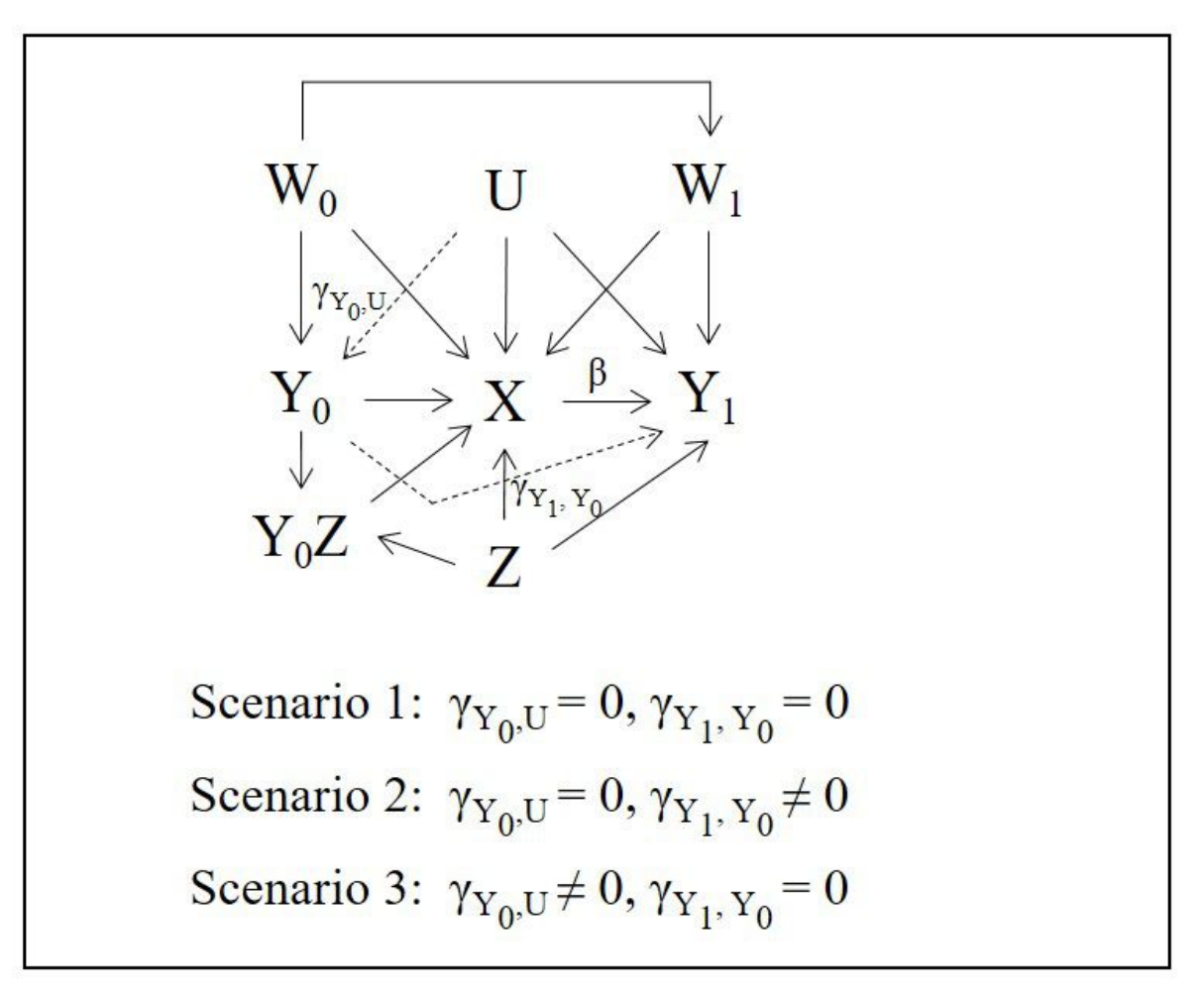}
	\caption{{Causal DAG consistent with the data generation of the simulation outlined in Section \ref{sec:sim2}}}
	\label{fig:DAG_sim2}
\end{figure}

For the simulation demonstrating the POA-IV and POA-CF estimates the data was generated using the same strategy as for the simulation explained in Section \ref{sec:sim2}, except for $X$ and $Y_1$. The data generation models are shown below and Figure \ref{fig:DAG_sim2}  shows the DAG explaining the mechanisms with which the simulation scenarios are implemented.   

\begin{table}[h]
\centering
\begin{tabular}{rcl}
$\beta$& = & 0.1    \\
$Z_{ij}$     & $\sim$ & $Bern(0.5)$ \\ 
$W_{0,i}$ & $\sim$ & $N(0,1)$ \\ 
$W_{1,i}$ & $=$ & $\gamma_{W_1,W_0} W_{0,i} + \gamma_{W_1,\varepsilon}\varepsilon_{W_1,i} $ \\ 
$\varepsilon_{W_1,i}$ & $\sim$ & $N(0,1)$ \\
$U_i$ & $\sim$ & $N(0,1)$ \\ 
$Y_{0,i}$ & $\sim$ & $Bern(\gamma_{Y_0,0} + \gamma_{Y_0, U} U_i + \gamma_{Y_0, W_0}W_{0,i}$) \\ 
$X_i$     & $\sim$ & $Bern(\gamma_{X,0} + \gamma_{X,Z}Z_{ij} + \gamma_{X,U}U_i + \gamma_{X,W_0}W_{0,i} + \gamma_{X,W_1}W_{1,i} + $ \\
& & \textcolor{white}{ooooo}$\gamma_{X,Y_0} Y_{0,i} + \gamma_{X,Y_0Z} \cdot Z_{ij} \cdot Y_{0,i})$\\
$Y_{1,i}$ & $\sim$ & $Bern(\gamma_{Y_1,0} + \gamma_{Y_1,U}U_i + \beta X_i + \gamma_{Y_1,W_1}W_{1,i} + \gamma_{Y_1,Z}Z_{ij} + \gamma_{Y_1,Y_0}Y_{0,i})$ \\ 
\end{tabular}
\end{table} 

\newpage

\section*{Appendix 5}\label{sec:appendix5}

The Monte Carlo standard errors (MCSE) are calculated based on Morris et al. 2019 \cite{Morris19} for the simulation presented in Section \ref{sec:sim2}.  The results are given in the table below for the performance measures: bias, mean squared error, coverage and type 1 error. 
\begin{longtable}{l|lllllll}
\hline
\multicolumn{1}{c|}{  } &   & CaT     & \multicolumn{1}{c}{IV} & \multicolumn{1}{c}{CF}  & \multicolumn{1}{c}{DiD} & \multicolumn{1}{c}{POA-IV} & \multicolumn{1}{c}{POA-CF}\\ \hline
\endhead
\multirow{5}{*}{Scenario 1}    & MCSE(bias) & 0.0615 & 0.1287 & 0.1288 & 0.1012 & 0.1496 & 0.1356 \\ 
 & MCSE(MSE) & 0.0047 & 0.1095 & 0.113 & 0.1809 & 0.0098 & 0.0083 \\ 
 & MCSE(coverage) & 1.5787 & 0.1996 & 0.1996 & 0 & 0.7271 & 0.676 \\ 
 & MCSE(T1E) & 1.5712 & 0.2442 & 0.2442 & 0 & 0.7451 & 0.7451 \\  \hline
\multirow{5}{*}{Scenario 2}    & MCSE(bias) & 0.0613 & 0.1353 & 0.1354 & 0.0974 & 0.1505 & 0.1502 \\ 
 & MCSE(MSE) & 0.0047 & 0.1112 & 0.1162 & 0.1559 & 0.0104 & 0.0105 \\ 
 & MCSE(coverage) & 1.5770 & 0.2442 & 0.2230 & 0 & 0.6415 & 0.6486 \\ 
 & MCSE(T1E) & 1.5753 & 0.2636 & 0.2442 & 0 & 0.6197 & 0.6343 \\    \hline
\multirow{5}{*}{Scenario 3}    & MCSE(bias) & 0.0629 & 0.1313 & 0.1314 & 0.0998 & 0.1493 & 0.1306 \\ 
 & MCSE(MSE) & 0.005 & 0.1142 & 0.1181 & 0.2327 & 0.0098 & 0.0079 \\ 
 & MCSE(coverage) & 1.581 & 0.1729 & 0.1729 & 0 & 0.6556 & 0.8174 \\ 
 & MCSE(T1E) & 1.5808 & 0.1729 & 0.1729 & 0 & 0.751 & 0.7626 \\      
\caption{Monte Carlo standard errors (MCSE) of the performance measures of all estimates and all scenarios of the simulation outlined in Section \ref{sec:sim2}. All results are multiplied with 100 and rounded to 3 significant figures. }\label{tab:MCSE_results_sim2}
\end{longtable}

\newpage
\section*{Appendix 6}\label{sec:appendix6}
\begin{figure}[!htbp]
	\centering
	\includegraphics[width=0.8\linewidth]{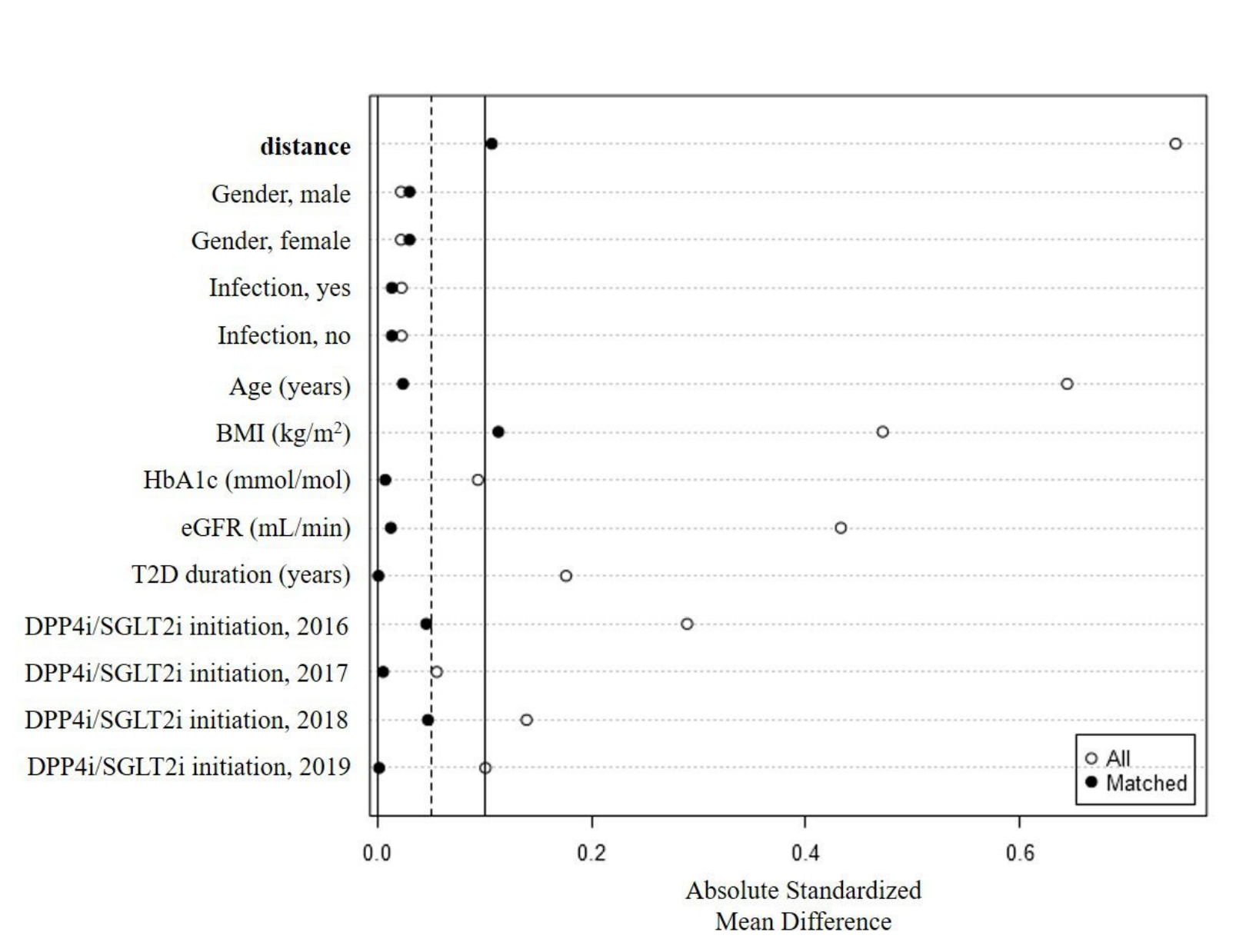}
	\caption{Love plot of the original and propensity score matched data.}
	\label{fig:love_plot}
\end{figure}

The propensity score matching procedure matched 100\% of the $1966$ individuals treated with SGLT2i. Therefore, overall 67.43\% of all individuals in the data where matched. No records were discarded for the matching procedure. The love plot in Figure \ref{fig:love_plot} shows that the matched data improved the balance of groups based on the absolute standardize mean difference. The matching process was employed using the baseline characteristics shown in the figure measured at first-line treatment initiation and years of second-line treatment initiation as this covariate has no effect on the treatment effect. 

\newpage
\section*{Appendix 7}\label{sec:appendix7}
Correlation plot shows the pairwise correlation of all estimates  using 500 bootstrap samples as explained in Section \ref{sec:application}. Estimates of the CaT and PSM as well as the estimates of the POA-IV and POA-CF are highly correlated. 

\begin{figure}[!htbp]
	\centering
	\includegraphics[width=0.8\linewidth]{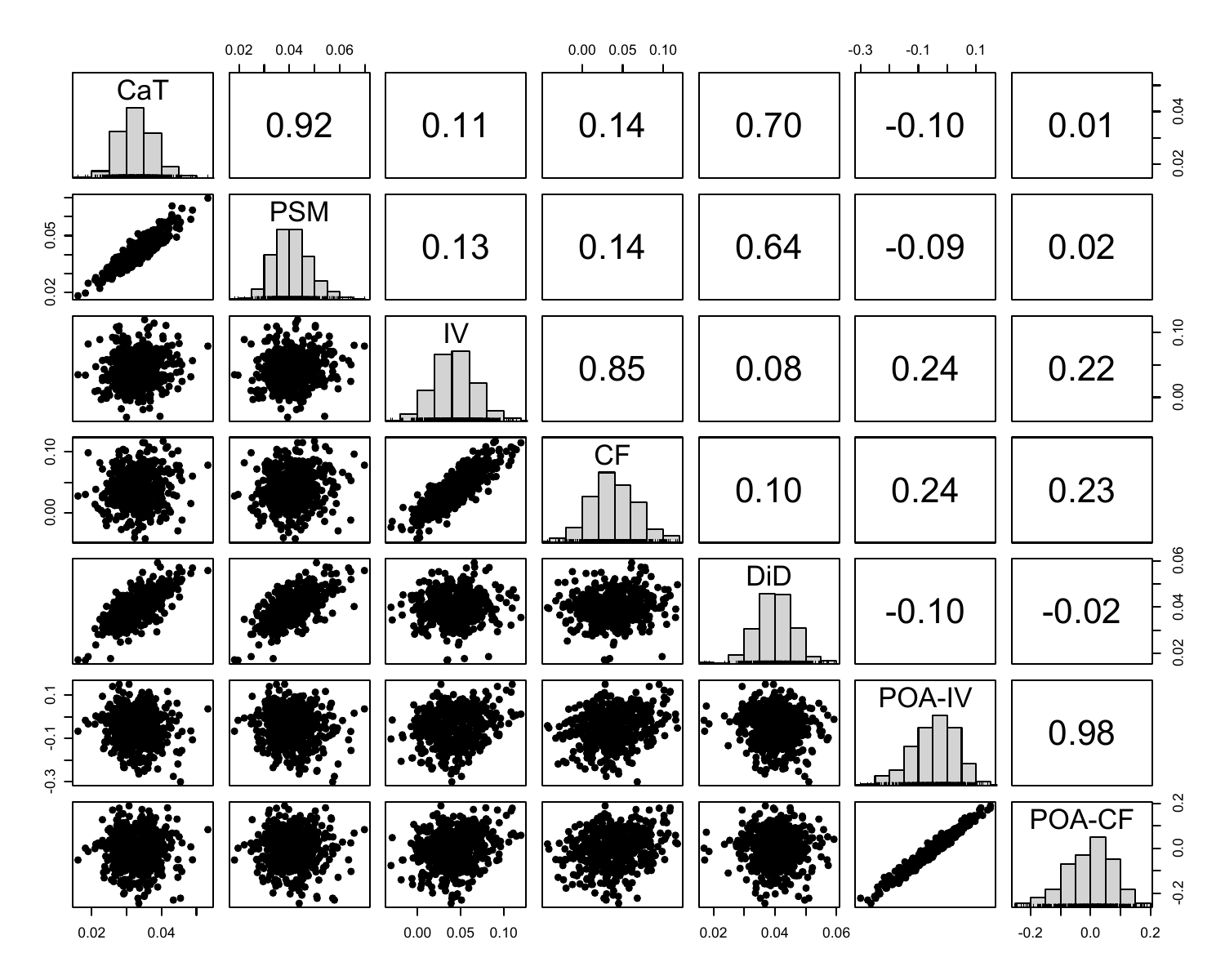}
	\caption{Correlation plot of all bootstrapped estimates of the application study.}
	\label{fig:correlation_plot}
\end{figure}

\end{document}